\documentclass[aps,twocolumn,showpacs,superscriptaddress,groupedaddress]{revtex4-1}  

\usepackage{graphicx}  
\usepackage{dcolumn}   
\usepackage{bm}        
\usepackage{amssymb}  
\usepackage{amsmath}
\usepackage{eqnarray,amsmath}
\usepackage{lipsum} 
\usepackage[colorinlistoftodos]{todonotes}
\usepackage[colorlinks=true, allcolors=blue]{hyperref}
\usepackage{physics}
\usepackage{float}
\usepackage{braket}
\usepackage{hyperref}
\usepackage{color}
\newcommand{\avg}[1]{\left<#1\right>}

\begin{document}

\title{Berry-electrodynamics - Anomalous drift and pumping from time-dependent Berry connection}
\author{Swati Chaudhary\textsuperscript{1}}
\email{swatich@caltech.edu}
\author{Manuel Endres\textsuperscript{1}}
\author{Gil Refael\textsuperscript{1}}
\affiliation{\textsuperscript{1}Institute of Quantum Information and Matter, California Institute of Technology, Pasadena, California 91125, USA}
\begin{abstract}
The Berry curvature of a Bloch band can be interpreted as a local magnetic field in reciprocal space. This analogy can be extended by defining an electric field analog in reciprocal space which arises from the time-dependent Berry connection. We explore the term in the semi-classical equation of motion that gives rise to this phenomenon, and show that it can lead to anomalous drift in wave packet motion. A similar effect arises from changes in the band population due to periodic driving, where the resulting drift depends on the nature of the drive and can be expressed in terms of a shift vector. Finally, these effects can be combined to build a pump with a net anomalous drift during a cyclic evolution in momentum space.
\end{abstract}
\maketitle

\section{Introduction}

The non-trivial geometry of energy bands in lattice models often gives rise to non-zero Berry curvature, which can lead to Hall response and affect material properties significantly~\cite{berry1984quantal,xiao2010berry,RevModPhys.82.3045, zhang2005experimental,price2014quantum}. Berry curvature can be interpreted as a local magnetic field in momentum space. Its effect on the semiclassical dynamics is well studied \cite{price2012mapping,diener2003intrinsic}. The local nature is directly observed in many cold atom setups, where localized wave packets in momentum space can be generated and coherently controlled~\cite{wunsch2008dirac,grusdt2014measuring,jotzu2014experimental,aidelsburger2015measuring}. In most of these experiments, the underlying band topology is revealed either by Aharanov-Bohm~\cite{aharonov1959significance} effects in quasimomentum space~\cite{duca2015aharonov,li2016bloch}, or by Hall drift measurements~\cite{jotzu2014experimental,dauphin2013extracting}, while new methods even allow the reconstruction of the Berry curvature across the Brillouin zone (BZ)~\cite{flaschner2016experimental,li2016bloch}. Most of these methods exploit the analogy between the Berry curvature and a magnetic field, and measure its effect on the dynamics of a wavepacket moving adiabatically in one of the Bloch bands. 
 
It is  natural  to ask what happens if we keep the wavepacket stationary in the BZ, but change the band geometry. What effects arise from a time-dependent band geometry as experienced by a wavepacket localized at some quasimomentum $\textbf{q}$? This situation can be realized either by making the band geometry time dependent, or by changing the relative band population of two bands with opposite geometric properties. We recount how the rate of change of the Berry connection appears as an electric field analog in the semiclassical equation of motion for a wavepacket undergoing adiabatic evolution~\footnote{This term is also mentioned in the review \cite{xiao2010berry} but its effects on transport were not explored or used as far as we know}. We connect this effect to the shift that a wavepacket undergoing Rabi oscillations between two bands with opposite geometric properties exhibits. 

Our results extend earlier results for thought experiments involving electrons moving slowly in spatially varying magnetic fields~\cite{stern1992berry}, where the time dependence of a Berry flux  gives rise to an analog of electromotive force and an associated motion.  In addition, non-trivial band geometry can profoundly affect the non-linear optical responses of a solid~\cite{morimoto2016topological,doi:10.1002/adma.201603345,wu2017giant,PhysRevLett.116.237402}. Inspired by the role of  band geometry in these non-adiabatic processes, we also explore the consequences of a  time varying average Berry connection arising due to excitations between bands.

Our main motivation is to explain how the Berry connection dynamics, both adiabatic and non-adiabatic could be used to control the motion of wavepackets. This is timely given the variety of experiments, particularly in the atomic and optical realm~\cite{li2016bloch,tarruell2012creating,duca2015aharonov,jotzu2014experimental,wimmer2017experimental}, which explore the motion of wavepackets rather than the transport properties of  a whole Fermi sea, as is typical in solid-state systems. In our work, we explore the anomalous motion that such Berry-dynamics produces for a wave packet in a honeycomb lattice. Furthermore, we show how these processes can be combined to produce deterministic translations of a wave packet, including a pumping cycle. Given that the motion is due to geometric effects, it has the advantage that it is by and large detail-independent. 

\section{ background and summary of results}

In order to understand the effects of a time-dependent band geometry, let us first  review the effects of Berry curvature on the motion of a wavepacket in a Bloch band. The Berry curvature effects on the  center of mass (COM) motion  of a wavepacket in the $n^{th}$ Bloch band are well captured by the  semiclassical equation of motion derived in Refs.~\cite{xiao2010berry,sundaram1999wave,Geometrodynamics}. For a wavepacket moving adiabatically in the $n^{th}$ band, the COM velocity becomes:

\begin{equation}
\textbf{v}^n(\textbf{q})=\nabla_\textbf{q}E^n(\textbf{q})+\dot{\textbf{q}}\times\mathbf{\Omega}^n,
\label{BCv2}
\end{equation}
where $E^n(\textbf{q})$ is the energy of the $n^{th}$ band, and $\mathbf{\Omega}^n(\textbf{q})$ is the Berry curvature given by 
\begin{equation}
\mathbf{\Omega}^n(\textbf{q})=\nabla\times\textbf{A}_{\text{nn}}, \quad\text{where}\quad
\textbf{A}_{\text{nn}}=\bra{u_n(\textbf{q})}i\nabla_\textbf{q}\ket{u_n(\textbf{q})}
\label{BC}
\end{equation}
is the Berry connection,  and $\ket{u_n(\textbf{q})}$ is the space periodic part of $n^{th}$ band eigenstate. This description shows that the COM velocity has a contribution from a Lorentz force analog in addition to the regular group velocity, and thus highlights the analogy between a magnetic field and the Berry curvature, in the sense of 
\begin{equation}
\textbf{B}\leftrightarrow\nabla\times\textbf{A}_{\text{nn}}=\mathbf{\Omega}^n. \nonumber
\end{equation}

In our work, we recount how in a more general scenario, the COM velocity is given by (see App. \ref{appendixa})
\begin{equation}
\label{EFv}
\textbf{v}^n(\textbf{q})=\nabla_{\textbf{q}}E^n(\textbf{q})+\dot{\textbf{q}}\times\mathbf{\Omega}^n+\left(\frac{\partial \textbf{A}_{\text{nn}}}{\partial t}\right)_\textbf{q}-\nabla_{\textbf{q}}\chi_n(t),
\end{equation}
where
\begin{equation}
\chi_n(t)=i\bra{u_n}\frac{\partial}{\partial t}\ket{u_n}.
\label{chi}
\end{equation}
The last two terms in Eq.~(\ref{EFv}) arise when the band structure is changed adiabatically. In the absence of a force, these terms can be treated as a correction due to the time dependence of the Berry connection, and hence as an analog of the electric field, in the sense of 
\begin{equation}
\mathbf{E}\leftrightarrow \frac{\partial}{\partial t}(\textbf{A}_{\text{nn}})-\nabla_{\textbf{q}}\chi_n(t). \nonumber
\end{equation}
Our extension to time-varying band-structures suggest an interpretation of $\textbf{A}_{\text{nn}}$ as vector potential and $\chi$ as the electric potential. Note that the term $\frac{\partial}{\partial t}(\textbf{A}_{\text{nn}})-\nabla_{\textbf{q}}\chi_n(t)$ is gauge invariant (see App.~\ref{appendixa}). We show that this additional term can give rise to an anomalous drift which is studied in Sec. \ref{Adiabatic}.  We note that a essentially the same term is also derived in the review \cite{xiao2010berry}, Eq. 6.9, for general changes in a band structure, its effect on transprot, however, have been so far unexplored. In particular, we see that the Berry connection is simply playing the role of a shift of the wavepacket center. 

These time-dependent Berry connection effects are band dependent. Intuitively, one would expect that the process of band switching (in a static band structure) should also be considered as an effective time-dependent change in the Berry-connection seen by a wavepacket, which could lead to similar results. Indeed, we show that for a wavepacket starting in one band and undergoing  Rabi oscillations between two bands with different geometric properties, the COM velocity is given by (see App.~\ref{appendixb}):
\begin{equation}
\textbf{v}=\avg{\nabla_{\textbf{q}}E_n}+\frac{\partial }{\partial t}\avg{\textbf{A}_{\text{nn}}}+\frac{\partial}{\partial t}\avg{\nabla_\textbf{q}(\phi_n)},
\label{non-adiabaticvone}
\end{equation}
where
\begin{equation}
\phi_1=-\phi_2=(\text{Arg}{\bra{u_1}H'\ket{u_2}})/2
\end{equation}
is the phase of the matrix element connecting the two bands via the perturbation Hamiltonian $H'$ inducing the Rabi oscillation. For any quantity $O$, we define the average $\avg{O_n}=P_1 O_1+P_2 O_2$ with $P_1$ and $P_2$ the occupation probabilities for the two bands. The first term in Eq.~(\ref{non-adiabaticvone}) is  the average group velocity, and the last two terms can be considered as an anomalous correction arising due to the change in the average Berry connection, and the $q$ dependence of the phase of the transition matrix element. In this case, the electric field analogy is
\begin{equation}
\mathbf{E}\leftrightarrow \frac{\partial}{\partial t}\avg{\textbf{A}_{nn}}+\frac{\partial}{\partial t}\avg{\nabla_\textbf{q}(\phi_n)}. \nonumber
\end{equation}
We show gauge independence in App.~\ref{appendixb}. Depending on the nature of the drive, which modifies $\frac{\partial}{\partial t}\avg{\nabla_\textbf{q}(\phi_n)}$, the electric field term can lead to an anomolous drift (Sec. \ref{nonadiabatic}).

Most importantly, we show in Sec. \ref{pump} how to construct a charge pump by combining and repeating adiabatic and non-adiabatic steps. An alternative scheme for a pump, combining non-adiabatic processes with and without anamolous drift, is presented in App. \ref{nonadiabaticpump}.

\section{Anomalous drift from adiabatic changes of the band structure}
\label{Adiabatic}

Consider a Hamiltonian $H(\textbf{q},\textbf{G}(t))$ which depends on quasimomentum $\textbf{q}$ and a set of time-dependent parameters denoted by $\textbf{G}(t)$.  When the parameters $\textbf{G}$ are varied in an adiabatic manner, the COM velocity of a wavepacket initialized in the $n^{th}$ Bloch band is given by:
\begin{equation}
\textbf{v}(q)=\nabla_{\textbf{q}}E_n(q)+i\frac{\partial G_{\mu}}{\partial t}\left[\left<\frac{\partial u_n}{\partial G_{\mu}}|\nabla_{\textbf{q}} u_n\right>-\left<\nabla_{\textbf{q}} u_n|\frac{\partial u_n}{\partial G_{\mu} }\right>\right]
\end{equation}
where $\ket{u_n}$ is the space-periodic part of Bloch wave function for $n^{th}$ band (see App. \ref{appendixa}). This reduces to Eq.~(\ref{EFv}) with $\dot{\textbf{q}}=0$, where the last two terms can be interpreted as an electric field analog.

We demonstrate this effect by studying the honeycomb lattice. We consider a wavepacket  localized at quasimomentum $\textbf{q}=q_0\hat{x}$ as measured from the nearest Dirac point. In the vicinity of a Dirac point, the  Bloch Hamiltonian for the lowest two bands in A-B basis is
\begin{equation}
H=
 \frac{3}{2}J \begin{bmatrix}
    \Delta(t)& \tau_z q_x+iq_y\\
    \tau_zq_x-iq_y &-\Delta(t)
  \end{bmatrix},
 \label{Hoo}
\end{equation}
where  $\tau_z=\pm1$ for the two Dirac points $\textbf{K}_{\pm}$,  and $\textbf{q}=\textbf{k}-\textbf{K}_{\pm}$~\cite{RevModPhys.81.109}. The sublattice offset-energy $\Delta(t)$, which can be a function of time $t$, is measured in units of $\frac{3J}{2}$, where $J$ is the hopping  amplitude, and quasi-momentum $\textbf{q}$ in  units of $\frac{1}{a}$, where $a$ is the lattice constant. 

Consider a wavepacket in the lower Bloch band, and localized at $\textbf{q}=q_0\hat{x}$ in the vicinity of a Dirac point with $\tau_z=1$. For a time-dependent sublattice offset-energy $\Delta(t)$, the Berry connection is given by
\begin{equation}
\label{vectorpotential}
\textbf{A}_{\text{gg}}=\frac{1}{2q_0}\left(\frac{\Delta}{\sqrt{\Delta^2+q_0^2}}-1\right)\hat{y},
\end{equation}
where the gauge is chosen such that $\nabla_\textbf{q}(\chi)=0$.

\begin{figure}
 \includegraphics[scale=0.28]{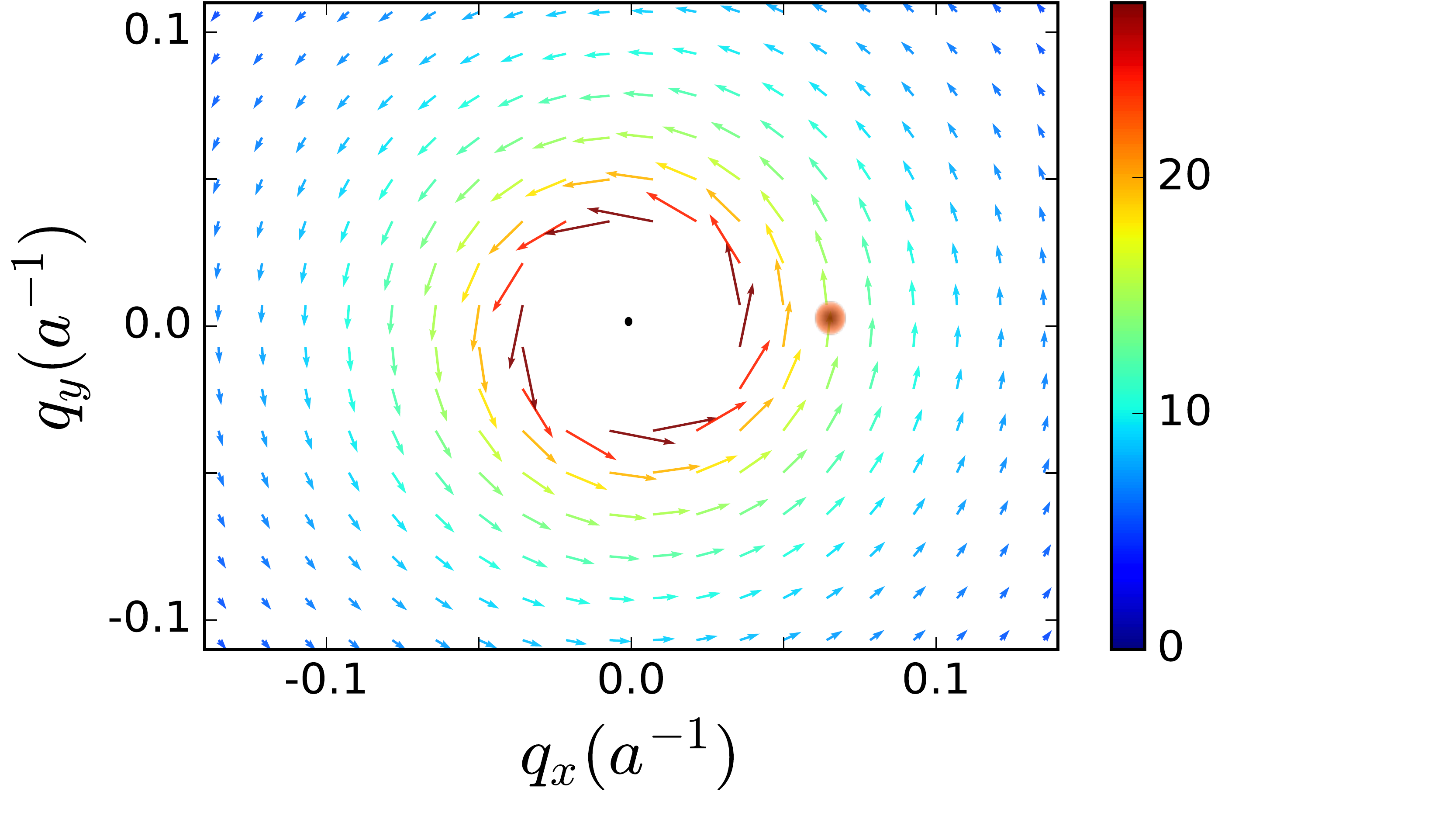}
 \caption{This vector plot shows the  difference in Berry connection (Eq.~(\ref{vectorpotential})) for the lower band eigenstate around a Dirac point  when the sublattice offset-energy is changed adiabatically from  $-\Delta_0$ to $\Delta_0$. The color bar on the side represents the magnitude of this difference. The  wavepacket position is marked by a red circle. Here, the gauge is chosen such that $\nabla_{\textbf{q}}\chi$ in Eq.~(\ref{anomvelocity}) vanishes and thus the anomalous drift after the adiabatic evolution is the same as the change in the Berry connection vector at the position of wavepacket. }
 \label{BCvectplot}
 \end{figure}

Changes of the Band structure are induced by varying  $\Delta(t)$ linearly from $-\Delta_0$ to $\Delta_0$ in time $T$. Then in the semiclassical picture, the group velocity and the anomalous velocity are given by:
\begin{equation}{\label{groupvelocity}}
\textbf{v}_g =\nabla_{\textbf{q}}E_- = -\frac{ \textbf{q}}{\sqrt{\Delta^2 +q_0^2}},
\end{equation} and 
\begin{equation}
 \textbf{v}_a=\frac{\partial \textbf{A}}{\partial t}-\nabla_\textbf{q}\chi_n(t)=\frac{\partial\Delta}{\partial t}\frac{q_0}{2(q_0^2+\Delta^2)^{3/2}}\hat{y}
 \label{anomvelocity}
 \end{equation}
 with $\Delta(t)=\frac{2\Delta_0}{T}(t-\frac{T}{2})$. When $\nabla_\textbf{q}(\chi)=0$,  the anomalous drift depends only on  the change in the Berry connection which is shown in Fig.~\ref{BCvectplot}, and it is significant only in the vicinity of a Dirac point.

The total anomalous displacement could be integrated, as it is simply:
\begin{equation}
\begin{split}
\delta_a=&\int\limits_{-\infty}^{\infty} dt \textbf{v}_a =A_{\text{gg}}(t=\infty)-A_{\text{gg}}(t=-\infty)\\=&\frac{1}{q_0}\frac{\Delta}{\sqrt{\Delta^2+q_0^2}}.
\end{split}
\end{equation}
This formula in particular establishes the Berry connection as simply a shift of the center of the wavepacket, and makes it significantly less abstract. 

\begin{figure}
\includegraphics[scale=0.29]{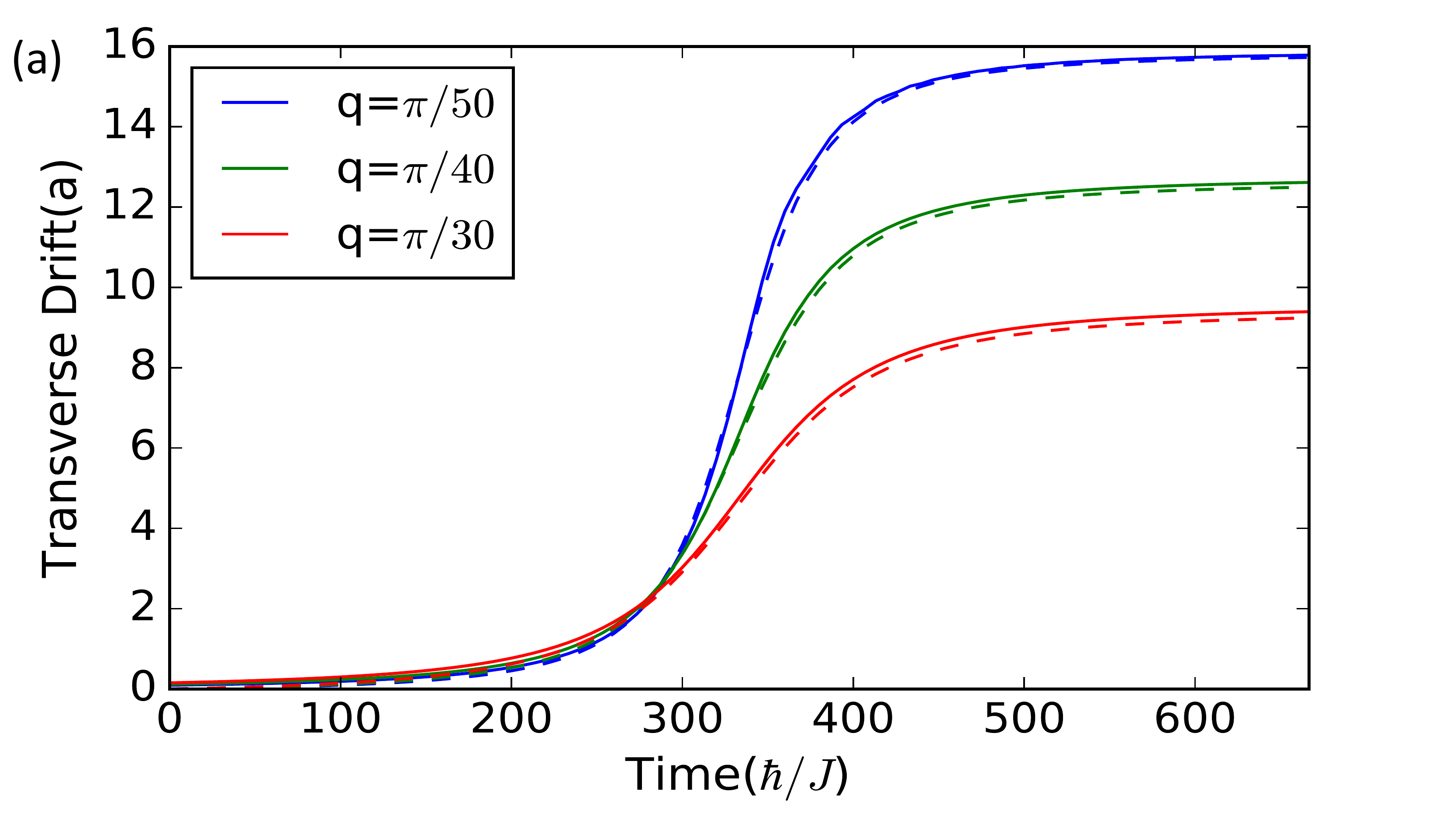}
\includegraphics[scale=0.29]{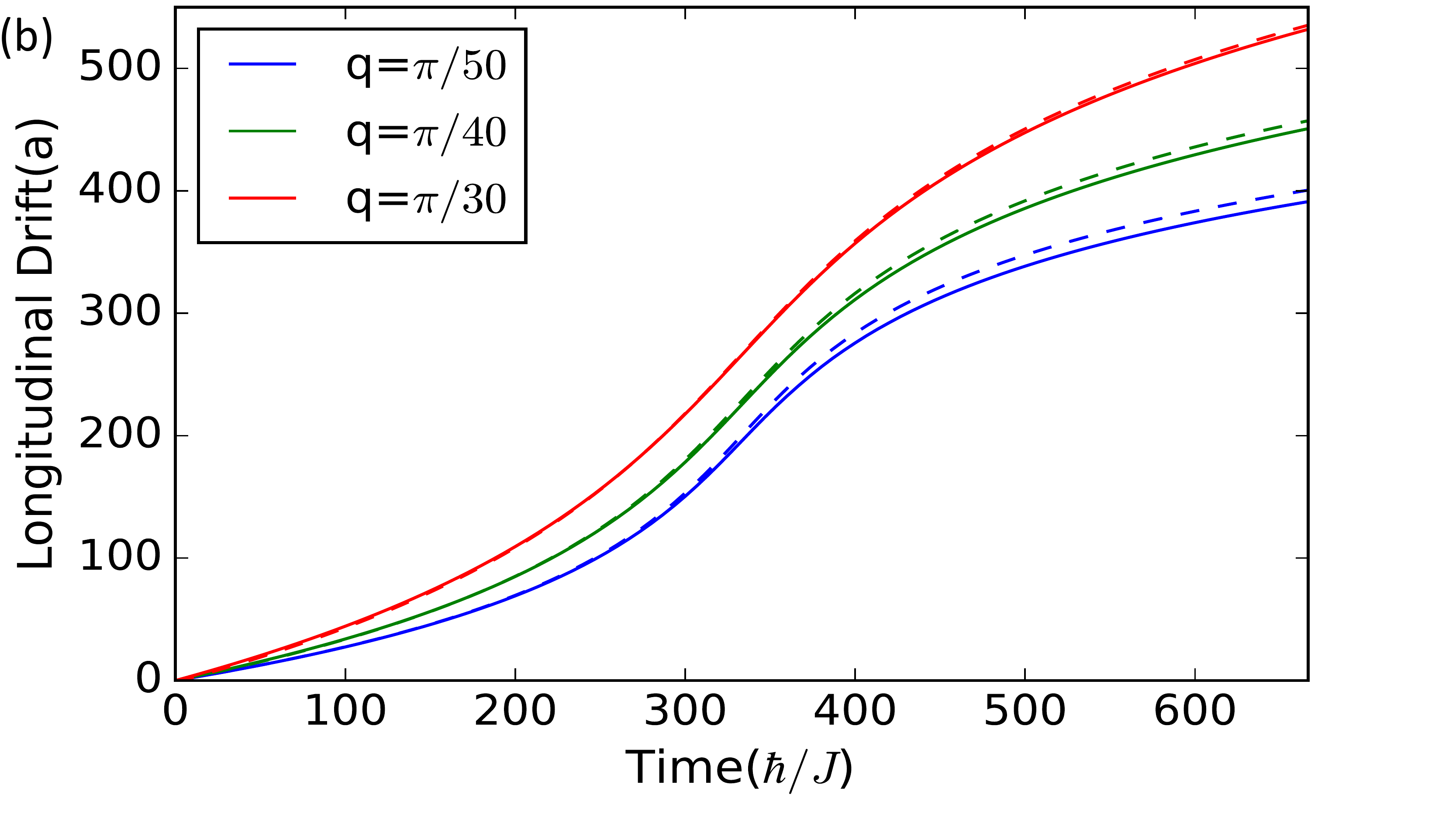}
\caption{COM displacement vs time when $\Delta$ is changed linearly in time from -0.4 to 0.4  (a) Transverse drift (units of a), (b) COM displacement  due to the group velocity term. Dashed lines depict the results from the semiclassical theory, and solid lines are from numerics.}
\label{TD}
\end{figure}

Adiabaticity is crucial for this result. Here, the adiabaticity condition is decided by Landau-Zener parameter, 
 $\Gamma=\frac{|v_{12}|^2}{\partial\mathcal{E}/\partial t}$, where $|v_{12}|$ is the gap at the level crossing, and $\mathcal{E}$ is the energy gap between two levels  far from the level crossing~\cite{PhysRevA.23.3107}. When $\Delta$ is changed linearly in time from a large negative to a large positive value, the energy gap $|v_{12}|=q$,  $\frac{\partial \mathcal{E}}{\partial t}=2\frac{\partial\Delta}{\partial t}$, and thus the Landau-Zener parameter $\Gamma=\frac{q^2}{\Delta_0/T}$. This process is adiabatic if $\Gamma>>1$.

To compare with the semi-classical expression, we numerically simulated the motion of a wavepacket centered at quasimomentum $q_0$ with spread $\sigma_q$,  in a honeycomb lattice  for the following set of parameters:  $q_0=\pi/50$, $\sigma_q=0.02$, T=700$\hbar/J$, $\Delta_0=0.4$. The  observed transverse drift, as shown in Fig.~\ref{TD}, is in good agreement with Eq.~(\ref{anomvelocity}). For the given set of parameters, the motion is almost adiabatic.  Indeed, for the numbers used we find at the center of the wavepacket, $\Gamma=2.59$, and excitation probability averaged over the gaussian wavepacket , $P_e\approx \left<e^{-2\pi\Gamma}\right>=0.0007$. Accordingly, we observe from numerics that the excitation probability $P_e\approx0.08\%$ for $q=\frac{\pi}{50}$.

\section{Anomalous drift from changes in the band population}
\label{nonadiabatic}

Non-adiabatic processes, involving bands with different geometry, can have various interesting consequences, e.g., effects originating from the shift in the charge center upon excitation~\cite{sipe2000second,young2012first,kral2000quantum,von1981theory,PhysRevLett.85.1512}. In many non centro-symmetric crystals, the difference between the Berry connection of the valence and conduction bands can give rise to a bulk photovoltaic effect during the optical transitions~\cite{fridkin2001bulk,PhysRevB.84.094115,PhysRevB.90.161409}. This kind of response can be expressed in terms of a shift vector~\cite{ von1981theory,sipe2000second}, which appears naturally in the study of shift current photovoltaic and photo galvanic effects~\cite{tan,PhysRevLett.85.1512}. This shift vector highlights the role of the band geometry in many
non-linear optical processes~\cite{kim2017shift,morimoto2016topological,PhysRevB.94.035117}. In these works, transitions are mainly induced by light, but in a more general scenario, one can consider any time-periodic perturbation which  changes the band population. As we show below, the shift vector depends not only on the Berry connection of the two bands, but also on the phase of the transition matrix elements. We illustrate this effect by discussing two types of band switching processes stemming from sub-lattice offset modulation or a sinusoidal force. We discuss the relation of our work to previous works ~\cite{sipe2000second,young2012first,kral2000quantum,von1981theory,PhysRevLett.85.1512,fridkin2001bulk,PhysRevB.84.094115,PhysRevB.90.161409,tan,kim2017shift,morimoto2016topological,PhysRevB.94.035117} at the end of the section.

The changes in the average Berry connection affect the motion of a wavepacket undergoing coherent interband Rabi oscillations. For a wavepacket localized in momentum space, and evolving under a time-dependent and space-periodic Hamiltonian, the wavefunction is:
\begin{equation}
\ket{\Psi(\textbf{r},t)}=\int d\textbf{q}\,\phi(q(t),q_0)\, e^{i\textbf{r}.\textbf{q}}\ket{\Phi(\textbf{q},t)}
\label{main Psi},
\end{equation}
where $\phi(\textbf{q}(t),q_0)$ is localized around $\textbf{q}_0$, and
\begin{equation}
\ket{\Phi(\textbf{q},t)}=A(t)\ket{g(\textbf{q})}+B(t)\ket{e(\textbf{q})}
\end{equation}
 is the superposition of the energy eigenstates $\ket{g(\textbf{q})}$ and $\ket{e(\textbf{q})}$ in the two bands. In the absence of an external force and for a translationally-invariant Hamiltonian,  $\phi(q(t),q_0)=\phi(q(t=0),q_0)$ can be taken as real, and the displacement in  real space is given by:
\begin{equation}
\avg{\textbf{r}}=\int_{BZ} d\textbf{q}|\phi(\textbf{q},\textbf{q}_0)|^2\bra{\Phi(\textbf{q},t)}i\nabla_{\textbf{q}}\ket{\Phi(\textbf{q},t)}.
\label{rr1}
\end{equation}
This expression is also valid in the presence of a weak and time-periodic force, but captures only the average displacement as shown in  App. \ref{appendixb}.
 We consider a perturbation $H'$  of the form:
\begin{multline}
H(\textbf{q},t)=H_0+H'\\=\begin{bmatrix}
    E_g(\textbf{q})& 0\\
    0 &E_e(\textbf{q})
  \end{bmatrix}+\begin{bmatrix}
    0& \frac{|V|}{2}e^{i\Theta}e^{i\omega t}\\
    \frac{|V|}{2}e^{-i\Theta}e^{-i\omega t} &0
  \end{bmatrix}
\label{RabiH}
\end{multline}
in the basis $\left\lbrace\ket{g(\textbf{q})},\ket{e(\textbf{q})}\right\rbrace$, where $\ket{g(\textbf{q})}$ and $\ket{e(\textbf{q})}$ are  lower and upper band energy eigenstates with eigenvalues $E_g$ and $E_e$, respectively. For the near resonance condition, $\omega\approx E_e(\textbf{q}_0)-E_g(\textbf{q}_0)$, it is useful to express $\ket{\Phi(\textbf{q},t)}$ as:
 \begin{equation}
\ket{\Phi(\textbf{q},t)}=a(t)e^{i\frac{\omega t}{2}}\ket{g(\textbf{q})}+b(t)e^{-i\frac{\omega t}{2}}\ket{e(\textbf{q})},
\label{adiabatic_phi}
\end{equation}
where $a(t)$ and $b(t)$ are slowly varying functions of time. Assuming that at $t=0$, a  wavepacket tightly localized at $\textbf{q}=\textbf{q}_0$ starts in the lower band, and if one ignores the terms oscillating at frequency $\omega$, the COM velocity as given by Eq.~(\ref{non-adiabaticvone}), now becomes
\begin{equation}
\textbf{v}=\textbf{v}_g+\textbf{v}_a,
\end{equation} 
where the average group velocity is
\begin{equation}
\textbf{v}_g=P_g\nabla_{\textbf{q}}E_g(\textbf{q})|_{\textbf{q}_0}+P_e\nabla_\textbf{q}E_e(\textbf{q})|_{\textbf{q}_0}=-\cos(2\Omega_{\text{eff}}t)\nabla_\textbf{q}E_0,
\label{groupvg}
\end{equation}
and the anomalous correction is given by
\begin{equation}
\begin{split}
\textbf{v}_a &= \frac{\partial}{\partial t}\left(P_g\textbf{A}_{\text{gg}}+P_e\textbf{A}_{\text{ee}}\right)- \frac{1}{2}\frac{\partial}{\partial t}\left(P_g\nabla_\textbf{q}\Theta-P_e\nabla_\textbf{q}(\Theta)\right)
\\&=\sin(2\Omega_{\text{eff}}t)\left(\textbf{A}_{\text{ee}}-\textbf{A}_{\text{gg}}-\nabla_\textbf{q}\Theta\right)\Omega_{\text{eff}}|_{\textbf{q}_0},
\label{nonadiabaticv}
\end{split}
\end{equation}
with phase $\phi_1=-\phi_2=\Theta/2$, defined in Eq.~(\ref{RabiH}), probability $P_e =1-P_g=|b(t)|^2=\sin^2(\Omega_{\text{eff}}t)$, and $\Omega_{\text{eff}}$ is the effective Rabi frequency.

\begin{figure}
\includegraphics[scale=0.5]{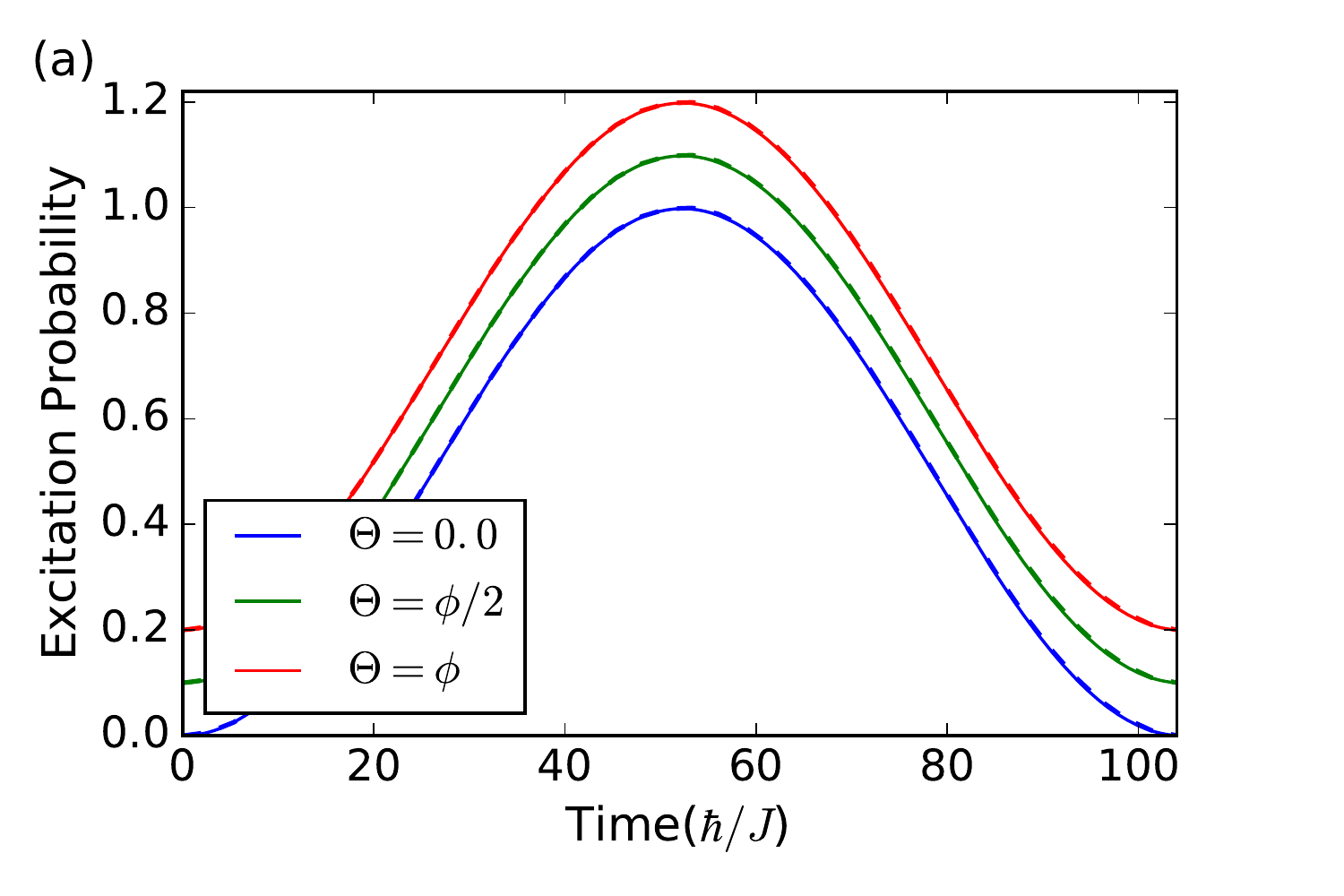}
\includegraphics[scale=0.5]{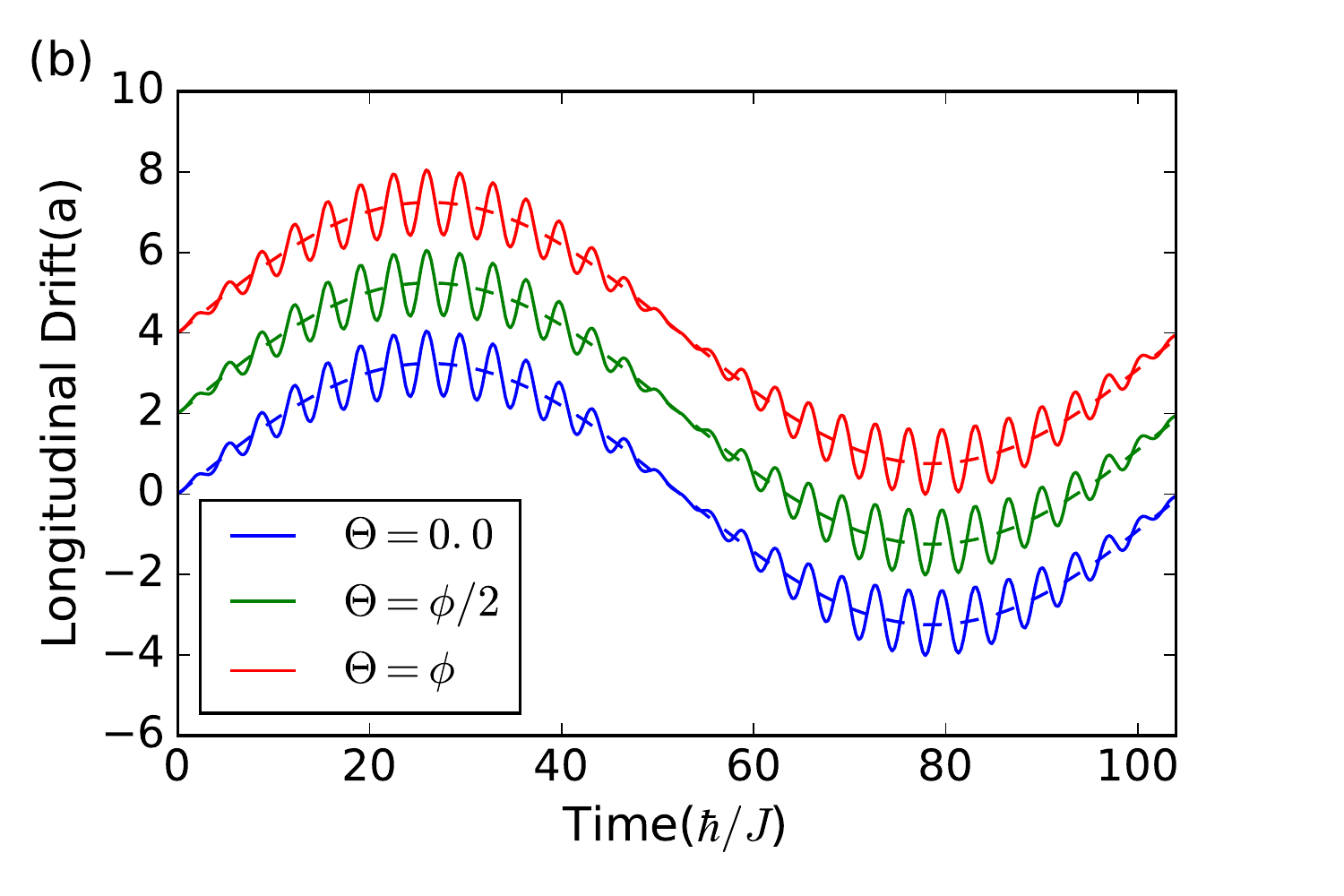} 
\includegraphics[scale=0.5]{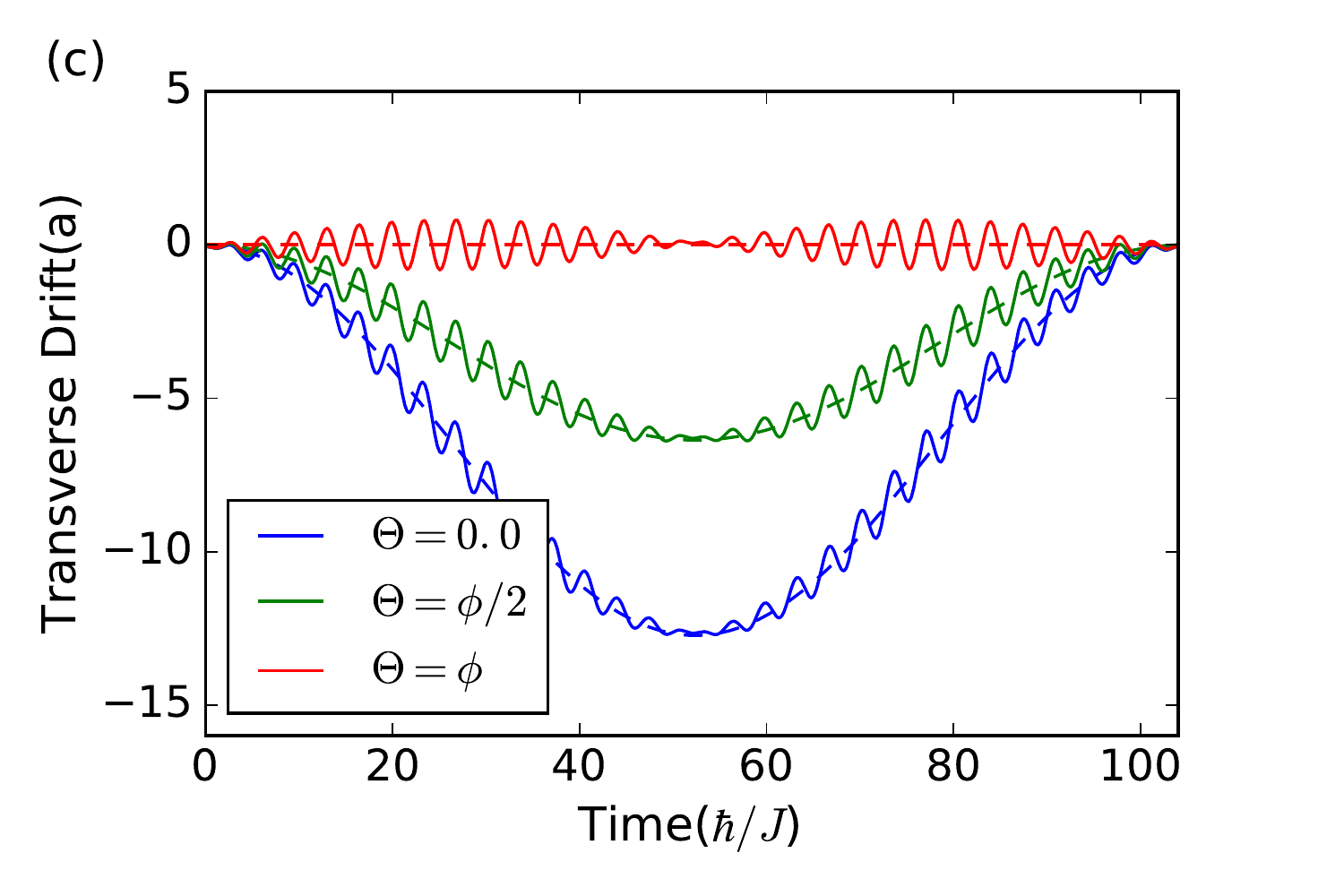}
\caption{(a) Transition probability as a function of time, (b)Longitudinal drift, (c) Transverse Drift for the time evolution by H (solid lines)in Eq.~(\ref{RabiH}) and from expressions(dashed lines) in Eq.~(\ref{groupvg}) and (\ref{nonadiabaticv}) respectively for $q_0=\pi/40$, $\Delta_0=0.6$, $V=0.04$, $\sigma_q=0.01$ and $\omega=2E_0$ and different  drive phases $\Theta$ where $\phi =\tan^{-1}(q_y/q_x)$. For the illustration purpose, the quantities on $y$ axis in (a) and (b) are shifted by a constant value for different $\Theta$. }
\label{Theta2}
\end{figure}

The anomalous velocity in Eq.~(\ref{nonadiabaticv}) depends on the difference in the Berry connection of the two bands, and the $\textbf{q}$ space gradient of the phase of the drive. We used a fixed basis to express $\ket{\Phi(\textbf{q},t)}$ in Eq.~(\ref{adiabatic_phi}), and both of these contributions are gauge-dependent, but the overall gauge dependence cancels. As a result, the anomalous velocity is proportional to a gauge-invariant quantity, $\textbf{A}_{\text{ee}}-\textbf{A}_{\text{gg}}-\nabla_\textbf{q}\Theta$, which is known as the shift vector in the context of non-linear optical processes~\cite{ sipe2000second}. Importantly, because of its dependence on the phase $\Theta$ of the drive, the resulting anomalous velocity $\textbf{v}_a$ can differ significantly, as shown in Fig.~\ref{Theta2}. 

Comparing the semi-classical result with the exact dynamics of a very narrow wavepacket (see Fig.~\ref{Theta2}), we notice that apart from the fast oscillations in COM motion, the dynamics is captured very well by Eq.~(\ref{non-adiabaticvone}). These fast oscillations arise from the non-zero inter-band Berry connection as shown in Eq.~(\ref{oscillations}) of App.~\ref{appendixb}. It is worth mentioning that for a large wavepacket  in quasimomentum space, finite detuning effects can cause significant deviation from the semi-classical theory, and the dependence on wavepacket size is discussed in App.~\ref{different_sigma}.

We now consider two different mechanisms to switch band population for a wavepacket in a honeycomb lattice, and  show how the nature of drive decides the anomalous shift during the transition.

\subsection{Band switching with $\Delta$ modulation} 
\begin{figure}
\includegraphics[scale=0.5]{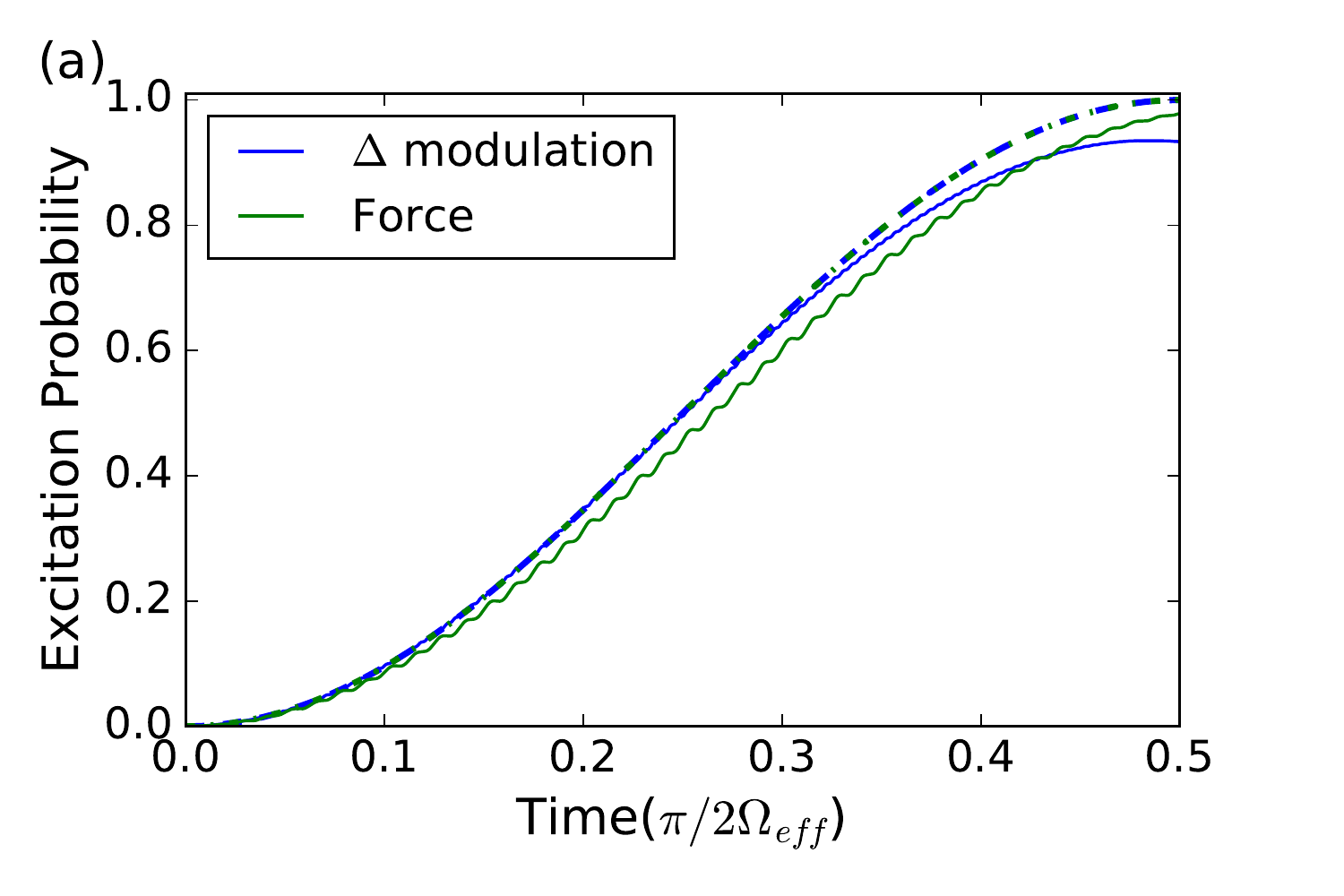}
\includegraphics[scale=0.5]{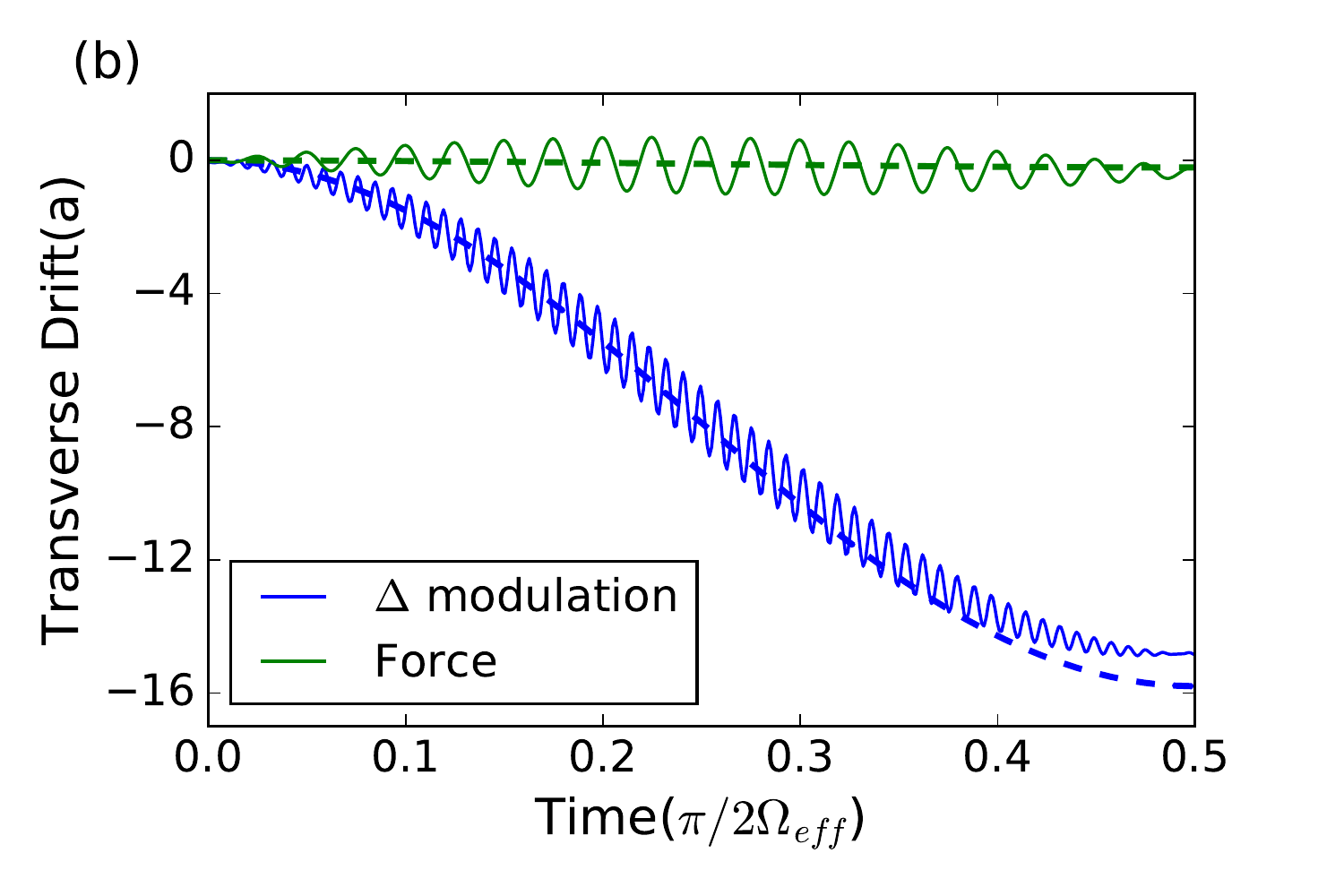}
\caption{(a)Transition probability and (b) Transverse drift as a function of time for two non-adiabatic schemes from numerics (solid line) and from theory (dashed lines). Results for $\Delta$ modulation are shown in blue color and we observe a significant drift during the band switching process as expected from Eq.~(\ref{delmodva}). On the other hand, when band switching is achieved by applying a sinusoidal force, anomalous drift (shown in green color) is vanishingly small and is in good agreement with Eq.~(\ref{Yforce}). }
\label{comparefig}
\end{figure}

We first consider switching the bands by modulating the sublattice offset-energy, and the Hamiltonian for this process is given by
\begin{equation}
\begin{split}
H&=H_0 +H '\\&=
  \begin{bmatrix}
    \Delta_0& qe^{-i\phi}\\
    qe^{i\phi} &-\Delta_0
  \end{bmatrix} +\begin{bmatrix}
  \Delta_0 a_m \cos\omega_r t&0\\
  0&-\Delta_0 a_m\cos\omega_r t\end{bmatrix}
  \label{delmodHAB}.
  \end{split}
  \end{equation}
in A-B Basis. Now, expressing this Hamiltonian in energy eigenstate basis of $H_0$, and comparing it to Eq.~(\ref{RabiH}), we find, for $a_m<<1$:

\begin{equation}
V=|V|e^{i\Theta}=\Delta_0a_m\frac{q_0}{\sqrt{\Delta_0^2+q_0^2}}
\label{Vfordelmod},
\end{equation}
when the lower and upper band eigenstates are expressed in the same gauge. Using the same gauge to find $\textbf{A}_{\text{ee}}$ and $\textbf{A}_{\text{gg}}$, we find
from Eq.~(\ref{nonadiabaticv}), that the anomalous velocity is given by:
\begin{equation}
\textbf{v}_a=-\sin(2\Omega_\text{eff} t)\left( \frac{1}{q_0}\frac{\Delta_0}{\sqrt{\Delta_0^2+q_0^2}}\right)\Omega_{\text{eff}}.
\label{delmodva}
\end{equation}
We simulate the dynamics of a very narrow wavepacket  for  $\sigma_q=0.01$, $a_m=0.14$, $q_0=\pi/50.0$, $\Delta_0=0.5$, and $\omega_r=2\sqrt{\Delta_0^2+q_0^2}$, and the observed transverse drift is in good agreement with Eq.~(\ref{delmodva}) as shown in Fig.~\ref{comparefig}.  In this case, the effective Rabi frequency, $\Omega_{\text{eff}}$ is very small, and the resonance condition is satisfied only at the center of the wavepacket. This results in a small discrepancy between numerics and theory, which can be attributed to the finite detuning for a fraction of the wavepacket. 

\begin{figure*}
	\includegraphics[width=\textwidth,scale=0.1]{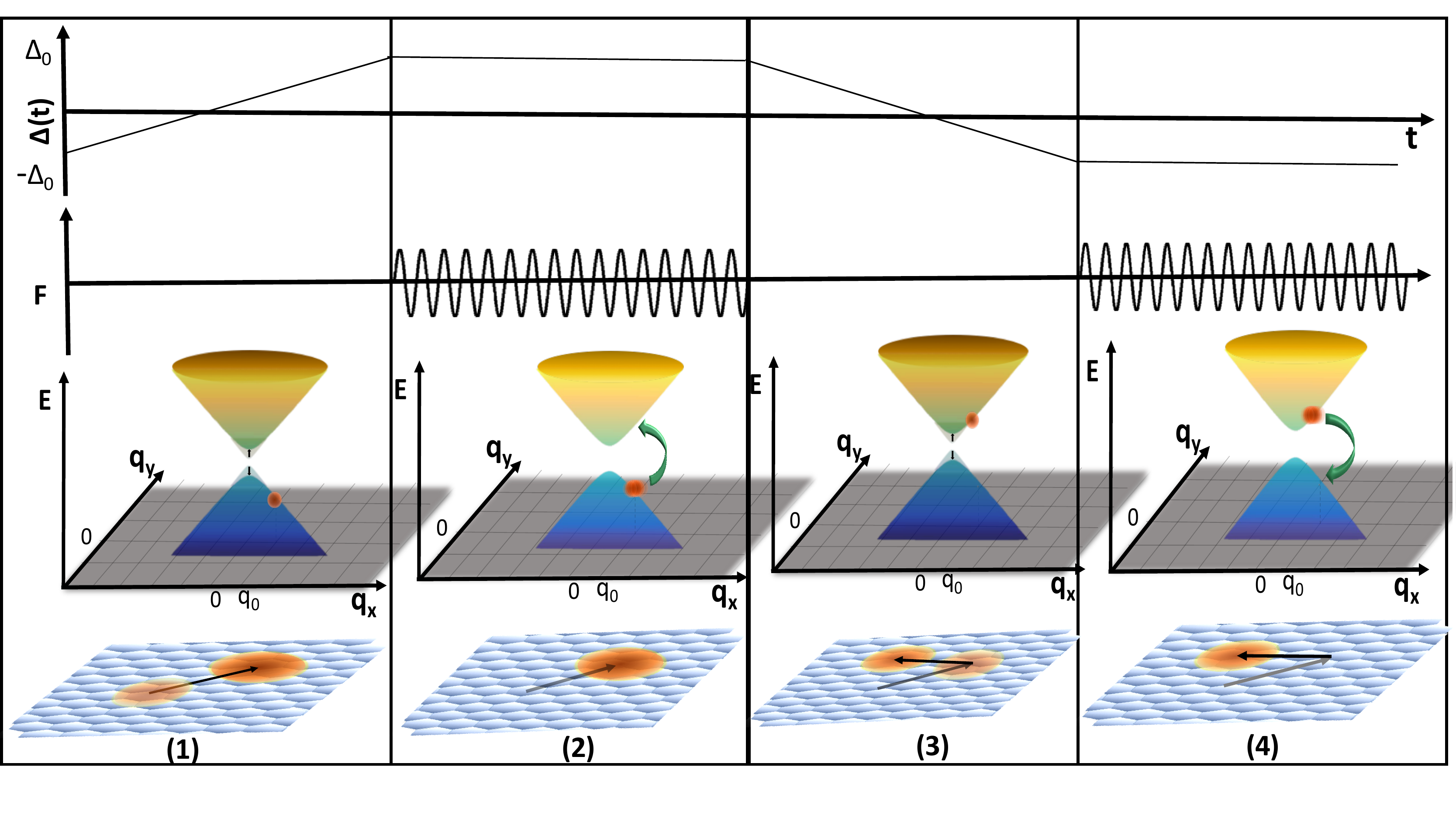}
	\caption{Scheme for amplification of the anomalous drift due to electric field analog. One cycle mainly consists of four steps (1) $\Delta$ is changed from $-\Delta_0$ to $\Delta_0$ adiabatically (2) $\Delta$ is  kept constant and a weak   sinusoidal force is applied to achieve band switching (3) $\Delta$ ramped adiabatically in the opposite direction (4) Again a $\pi$ pulse using a sinusoidal force. Upper two panels show how the sublattice offset-energy and external force is changed in time for different steps in the cycle. In third panel, we show the schematic for wavepacket localized in the vicinity of a Dirac point during the cycle. The lowermost panel illustrates the motion in real space. During each adiabatic step, the wavepacket center shifts significantly with transverse drift in the same direction. On the other hand, the time average displacement during the process of band switching is negligible. This results in an overall transverse drift during one cycle.}
	\label{scheme}
\end{figure*}

\subsection{Band switching with sinusoidal force} 
We further consider applying a time-periodic force on the wavepacket. We consider a weak external sinusoidal force, $\textbf{F}=\textbf{F}_0\sin\omega t$. If $|\textbf{F}_0\cdot\textbf{A}_{\text{ge}}|<<\omega$  (see Eq.~(\ref{F_H_appendixB}) in App. \ref{appendixb}  and Ref.~\cite{grusdt2014measuring}), the perturbation $H'$ is given by:
\begin{equation}
H' =\cos(\omega t)\begin{bmatrix}
\textbf{F}\cdot\textbf{A}_{\text{gg}}&\textbf{F}\cdot\textbf{A}_{\text{ge}}\\
\textbf{F}\cdot\textbf{A}_{\text{eg}}&\textbf{F}\cdot\textbf{A}_{\text{ee}}
\end{bmatrix}.
\label{F_H}
\end{equation}
Within rotating-wave approximation, we find from Eq.~(\ref{nonadiabaticv}) and~(\ref{F_H}) that  the anomalous velocity is given by:
\begin{equation}
\textbf{v}_a=\sin(2\Omega_\text{eff}t)\left(\textbf{A}_{\text{ee}}-\textbf{A}_{\text{gg}}-\nabla_{\textbf{q}}(\text{Arg}(\textbf{F}\cdot\textbf{A}_{\text{ge}}))\right)\Omega_{\text{eff}}.
\label{Yforce}
\end{equation}

Here, the anomalous term has two kind of contributions, one from the change in Berry connection, and the other from the phase of the drive. Interestingly, for the specific case of a wavepacket close to a Dirac point and  a linear force, these two contributions are equal. Hence,  the wavepacket is excited to the upper band without any significant transverse drift unlike the sublattice offset-energy modulation case (see Fig.~\ref{comparefig}). We also simulate the dynamics of a wavepacket for $\textbf{q}_0=\pi/40\hat{x}$, $\Delta_0=0.6$ in the presence of a sinusoidal force $\textbf{F}=\omega_r\Delta_0/20\cos(\omega_r t)\hat{y}$. The observed transverse drift and the transition probability are in good agreement with the analytical treatment (Fig.~\ref{comparefig}). The finite size of wavepacket results in some detuning effects as the resonance condition is fulfilled only the center of the wavepacket. This gives rise to a small discrepancy in numerics and theory. Also, we neglected the terms oscillating at frequency $\omega$ in Eq.~(\ref{oscillations}), and hence the oscillations in COM motion are not captured by Eq.~(\ref{nonadiabaticv}).

\subsection{Relation to previous shift-currents work}

During the band switching process, we came across interesting connections between the anomalous drift and the shift vector, $\textbf{A}_{\text{ee}}-\textbf{A}_{\text{gg}}-\nabla_\textbf{q}\Theta$. This shift vector also appears in the study of many other optoelectronic processes  in solids. It is responsible for shift current bulk photovoltaic effect~\cite{sipe2000second,young2012first,kral2000quantum,morimoto2016topological,PhysRevB.94.035117} and for the large nonlinear optical response in many non-centrosymmetric crystals. In these shift current calculations, one considers a completely filled valence band, and a continuous pumping to the conduction band gives rise to a shift current. The contribution from different $k$ points in BZ is proportional to the shift vector, and the proportionality constant depend on many other factors like amplitude and frequency of drive, temperature, etc. Since, the shift current is obtained by summing up this contribution from all $k$ points in a BZ, and thus it is non-zero only for crystals without an inversion center. Furthermore, the polarization of light directly affects the phase $\Theta$ of the transition matrix elements, and thus can change the shift vector significantly.

The effect for localized wavepackets depends only on the shift vector at the wavepacket center. Also, while electrons in crystals were treated through the Fermi golden rule, the wave packet we are considering exhibits coherent oscillations between the bands, and the observed anomalous drift oscillates in the same manner as the relative band population.  Most importantly, the transition matrix we consider is general, and does not apply only to optical transitions or nonlinear optical processes~\cite{sipe2000second,young2012first,kral2000quantum,morimoto2016topological,PhysRevB.94.035117}. In the context of the optoelectronic processes discussed in these references, the shift vector depends on the interband Berry connection, and is given by $\textbf{A}_{\text{ee}}-\textbf{A}_{\text{gg}}-\nabla_{\textbf{q}}(\text{Arg}(\textbf{F}\cdot\textbf{A}_{\text{ge}}))$. This dependence originates from the fact that the transition is induced by a time-periodic force, and thus according to Eq.~(\ref{F_H}),  $\Theta=\text{Arg}(\textbf{F}\cdot\textbf{A}_{\text{ge}})$.  Hence, our work generalizes the shift-current expression to $\textbf{A}_{\text{ee}}-\textbf{A}_{\text{gg}}-\nabla_\textbf{q}\Theta$, where $\Theta$ is the phase  of the matrix element connecting the two bands, and the previous results appear as a special case.

\section{Application - Pump from Electric field analog}
\label{pump}

Our primary goal is to show how time-dependent Berry connections could be used to control particle motion. 
Indeed, the two mechanisms to achieve an anomalous drift from time-dependent geometric properties as discussed above can be combined to produce a pumping effect on a wavepacket. During the adiabatic evolution (Sec.~\ref{Adiabatic}), the sign of transverse drift depended on the band index, and the rate of change of $\Delta$.  For the non-adiabatic case, we mentioned two schemes for band transition, and it was observed that the wavepacket displacement was negligible when a linear and time-periodic force was applied. These effects can be combined  to produce an amplified transverse drift (Fig.~\ref{scheme}), with the following steps:

\begin{enumerate}
\item{The wavepacket is initialized in the lower band, and the sublattice offset-energy $\Delta$ is changed adiabatically from -$\Delta_0$ to $\Delta_0$. This results in a transverse drift.}
\item{A $\pi$ pulse is realized by applying a weak external sinusoidal force  on the wavepacket. The wavepacket ends up in the upper band after this pulse with the same $q$ distribution as before. There is no significant transverse drift during this step. }
\item{Now, $\Delta$ is changed in the opposite direction. The transverse drift during this step is expected to be the same as in the step one because both, the band index, and the rate of change of $\Delta$ have opposite sign. }
\item{Another $\pi$ pulse is applied, and  the wavepacket returns to the lower band.}

\end{enumerate}
After these four steps, the wavepacket shifts by a finite distance in the transverse direction but returns back to the ground band with the same $q$-distribution. Importantly, also the lattice parameters are back to their initial values, which makes it possible to repeat these steps in a pump cycle. We discuss an alternative scheme only involving band transfer (but no adiabatic changes) in App.~\ref{nonadiabaticpump}.
 
The overall transverse drift after one complete cycle can also be associated with the total phase picked up by the state in this non-adiabatic but cyclic process.  We notice  that after  a full cycle of  the scheme shown in Fig.~\ref{scheme}, the space-periodic part of Bloch wavefunction in Eq.~(\ref{main Psi}) changes  from  $\ket{\Phi(\textbf{q},t)}=\ket{g(\textbf{q})}$ at $t=0$ to $\ket{\Phi(\textbf{q},t)}=e^{i\theta(\textbf{q})}\ket{g(\textbf{q})}$, where $\theta(\textbf{q})$ has contribution both from  dynamical and geometrical terms. Interestingly, in the scheme described above, the overall dynamical phase vanishes, and thus the phase picked up in one cycle can be interpreted as Aharanov-Anandan phase~\cite{AA}. Now, according to Eq.~(\ref{rr1}), the COM displacement is given by:
\begin{equation}
\avg{\textbf{r}}=\nabla_{\textbf{q}}\theta(\textbf{q})|_{\textbf{q}_0},
\label{rtotalphase}
\end{equation} for an extremely  narrow wavepacket located at  $\textbf{q}=\textbf{q}_0$. Here, the transverse after one cycle, depends solely on Aharanov-Anandan phase, and is thus non-zero only for those cyclic processes where the state picks up a non-trivial geometric phase.

We  explained above how this phase depends on the  nature of  $H'$, and  there is no $\mathbf{q}$ dependent  overall phase when band switching is obtained by  modulating the sublattice offset-energy.  On the  other hand, for the weak sinusoidal  force case, in the limit $|\Delta|>>q_0$, and for the same gauge choice,
\begin{equation}
\theta(\textbf{q})\approx\text{Arg}(\textbf{F}\cdot\textbf{A}_{\text{ge}}(\Delta_0))+\text{Arg}(\textbf{F}\cdot\textbf{A}_{\text{eg}}(-\Delta_0))\approx2\phi,
\end{equation}
where $\phi=\tan^{-1}(q_y/q_x)$,
and thus we expect the displacement after one cycle to be $<\textbf{r}>=2/q_0\hat{y}$.

\begin{figure}[h]
\includegraphics[scale=0.27]{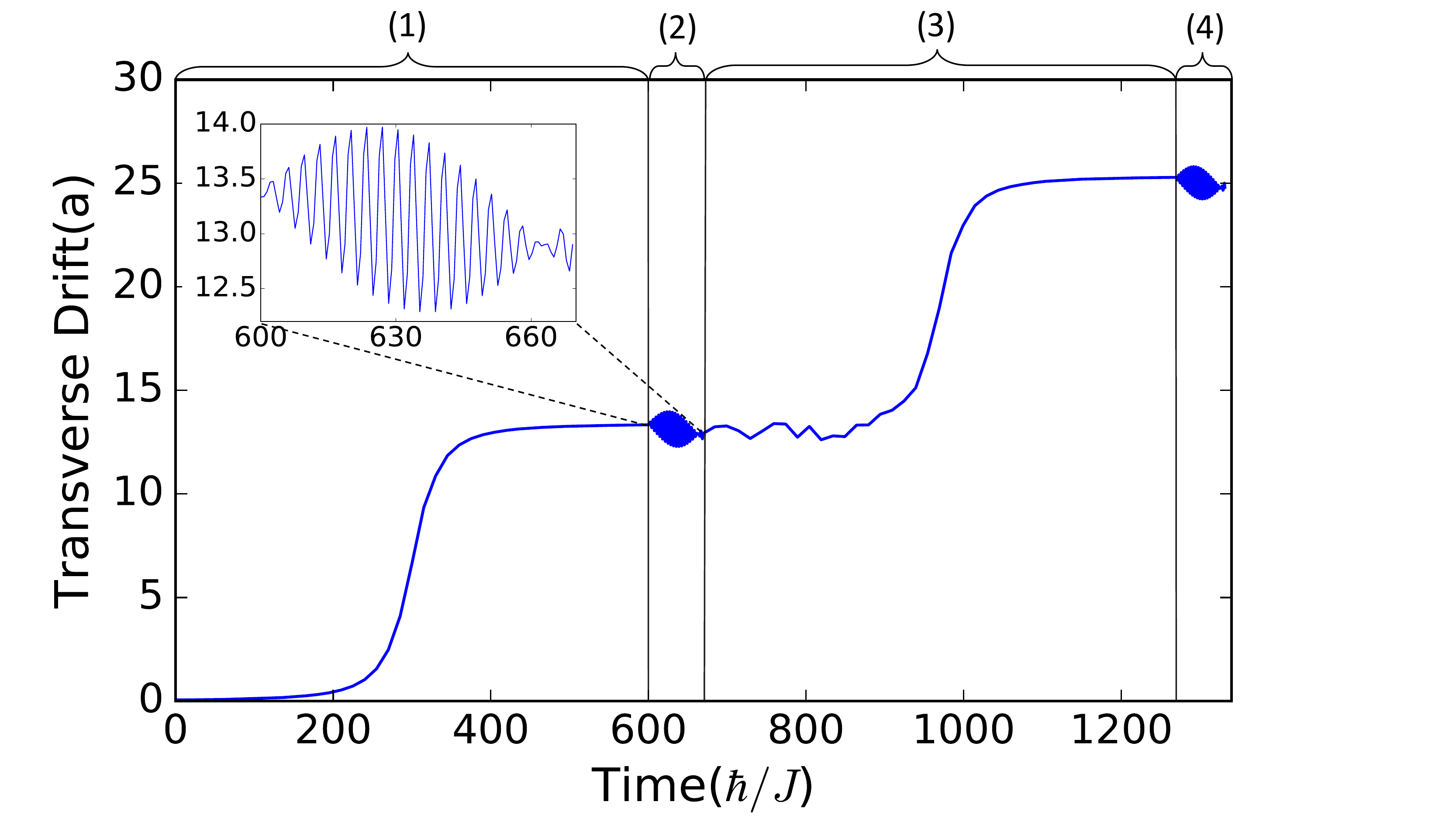}
\caption{Transverse drift for one cycle in scheme shown in Fig.~\ref{scheme}. 
	 We observe a transverse drift during the adiabatic evolution when the sublattice offset is varied from $-\Delta_0$ to $\Delta_0$ in step 1.  Next, the wavepacket is excited to the upper band without any significant drift when a sinusoidal force is applied during the step 2.
	  In step 3, it is displaced in the same direction during the adiabatic evolution  in the opposite band when sublattice offset is changed in the opposite direction. Again, the wavepacket returns to the lower band without any insignificant displacement during step 4.  In this case, the offset was changed very slowly from -0.6 to 0.6 in $T=600$ to satisfy the adiabaticity condition. The schematics of wavepacket dynamics in $q$ space is shown in Fig.~\ref{scheme}. }
\label{Amplify}
\end{figure}

We simulate the motion of wavepacket for one full cycle in Fig.~\ref{scheme} (see Fig.~\ref{twocycle} in Appendix for multiple cycles), and observe that the obtained transverse drift shown in Fig.~\ref{Amplify} is in close agreement with Eq.~(\ref{rtotalphase}). In this case, it was also observed that the wavepacket first expands in real space during the first adiabatic step, and then contracts during the second adiabatic step. It is mainly because the dispersion relation is opposite for two bands, and thus the wavepacket shows a breathing behavior. Hence, at the end of each cycle, we can achieve a significant transverse drift without any spreading of wavepacket in real space. 

\section{Discussion and Conclusions}

In this manuscript, we explored the time dependence of the Berry connection as a means of controlling a wavepacket in a Bloch band. Indeed, there are many ways to control a particle moving in a confined potential. It could be subject to a force, or it could be subject to a variety of quench protocols. A change of a geometric quantity such as the Berry connection, however, is more likely to result in a universal response. Furthermore, we demonstrated that the time derivative of the Berry connection appears as an anomalous velocity  in the semiclassical equation of motion for a wavepacket, and can be considered as a momentum space analog of an electric field.

In addition, when a wavepacket is excited to a different band, another interesting connection emerges between the observed anomalous drift and the shift vector, see Eq.~(\ref{nonadiabaticv}). During the band switching, the wavepacket experiences an anomalous drift which  consists of the difference of the Berry connections of the two bands, and a $\textbf{q}$ space derivative of the argument of the  interband matrix element of the drive. And hence, we demonstrated that shift vector expression is more general than the one encountered in the light-induced transitions~\cite{sipe2000second,young2012first,kral2000quantum,von1981theory,PhysRevLett.85.1512,fridkin2001bulk,PhysRevB.84.094115,PhysRevB.90.161409,tan,kim2017shift,morimoto2016topological,PhysRevB.94.035117}, and explored it in two types of drive.

Indeed, the anomalous drift from the electric field analog should be thought of as a powerful element in the arsenal for controlling a wavepacket in synthetic systems such as optical lattices. Such effects often average out in solid state systems, since all transport effects are averaged over a Fermi sea. Furthermore, changing the lattice geometry in materials on transport-relevant time scales is quite difficult. Atomic and optical systems, in contrast, allow the observation of the motion of tight wavepackets~\cite{Latticeaccel}, and can realize a variety of time dependent protocols which can explore the effects from time-dependent band geometry.

In our manuscript, we provide an example for what the combination of the anomalous velocity effects from the time varying band geometry and interband transition could achieve.  We construct a protocol which controllably `pumps' a wavepacket perpendicular to its group velocity. The observed transverse drift could be amplified without any significant change in the wavepacket size. 

Anomalous drifts from time-dependent Berry connection could be at the basis of a rich set of control protocols. It would be interesting to study the effects of time-dependent geometric properties for the degenerate Bloch bands or multi-band systems. Furthermore, geometric effects due to interband excitations could be made richer by considering a momentum shift $\textbf{q}$. Time-dependent band geometry could even have an effect in solid state system, perhaps when the Fermi sea is shifted due to a constant external field, or a thermal gradient. 
Beyond wavepacket dynamics, Berry curvature can also modify the energy spectrum of excitons, and can give rise to orbital Zeeman shift analog~\cite{PhysRevLett.115.166802,PhysRevLett.115.166803}. It would be interesting to explore stark shift analogs arising from time-dependent Berry connection.
Additionally, higher order moments of Berry curvature can also affect the magneto-optical properties of the materials~\cite{WSMPhysRevB.96.035205, WSPhysRevB.97.041101}. Using the recent proposals to control and tune the band geometry in monolayer materials~\cite{BCswitch,TunableBC11}, one can look up for the effects of time-dependent band geometry in many non linear optical responses as well. 
 We expect that these effects will be the subject of future investigation, as well as the focus of experimental efforts. 

\section{Acknowledgment}
We would like to thank Yuval Baum, Christopher D. White, and Karthik Seetharam for helpful discussions. We are grateful for support from the Institute of Quantum Information and Matter, an NSF Physics frontier center funded  by the Gordon and Betty Moore Foundation, and from the ARO MURI W911NF-16-1-0361 ``Quantum Materials by Design with Electromagnetic Excitation" sponsored by the U.S. Army.  SC is additionally grateful for support from the Barbara Groce Graduate Fellowship. GR gratefully acknowledges the hospitality of the Aspen Center for Physics funded by NSF PHY-1607611.

\newpage
\appendix
\section{Electric field analog in Equation of motion -Adiabatic Case }
\label{appendixa}
Consider a quantum system described by a Hamiltonian H which depends on quasi momentum $\textbf{q}$, and on a set of parameters given by $\textbf{G}$. The eigenstates of this $H$ are denoted by $\left |u_n(\textbf{q},\textbf{G})\right>$, such that,
\begin{equation}
	H(\textbf{q},\textbf{G})\left |u_n(\textbf{q},\textbf{G})\right>=  E_n(\textbf{q},\textbf{G}) \left |u_n(\textbf{q},\textbf{G})\right> .
\end{equation}
Now, consider the adiabatic evolution of this system as $\textbf{q}(t)$ and $\textbf{G}(t)$ changes slowly with time. According to the quantum adiabatic theorem, a system initially in one of its eigenstates $\left |u_n(\textbf{q}(0),\textbf{G}(0)\right>$ will remain in the instantaneous eigenstate of the Hamiltonian $H(\textbf{q}(t),\textbf{G}(t))$ throughout the process. The additional (geometric) phase picked by the state, when $\textbf{R}$ is varied slowly along the contour C is given by:
\begin{equation}
	\gamma _n =\oint_C d\textbf{R}\cdot\textbf{A}^n(\textbf{R}),
\end{equation}
where $ \textbf{R}= (\textbf{q},\textbf{G})$, and $\textbf{A}^n (\textbf{R}) $ is given by:
\begin{equation} \label{eq: Aq}
	\textbf{A}^n (\textbf{R})= i \left<u_n(\textbf{R})\right| \frac{\partial}{\partial\textbf{R}} \left|u_n(\textbf{R})\right>.
\end{equation}
In this case $\textbf{q}(t)$ is a 3D vector and we can take the dimension of $\textbf{G}$ as $m$, i.e, $\textbf{G}= (G_1,G_2,... G_m)$. We can define a gauge field tensor derived from the Berry vector potential:
\begin{equation} 
	\mathbf{\Omega^n_{\mu\nu}}= \frac{\partial}{\partial R_\mu}A^n_{R_\nu}-\frac{\partial}{\partial R_\nu}A^n_{R_\mu}, 
	\label{BC2}
\end{equation}
known as the Berry Curvature. In this case $\textbf{R}= (q_1,q_2,q_3, G_1, G_2, ...... G_m)$, and thus we can recast the Eq.~(\ref{BC2}) into a vector form for $\mu, \nu$ $  \epsilon$ $ (1,2,3)$
\begin{equation}\label{eq: vector form}
	\mathbf{\Omega}^n(\textbf{q})=\mathbf{\nabla}_\textbf{q}\times \textbf{A}^n(\textbf{q}),
\end{equation}
where the vector $\mathbf{\Omega}^n$ is related to the Berry curvature tensor by $\Omega^n_{ij}= \epsilon_{ijk}(\mathbf{\Omega}^n)_k$. The main point to be noted here is that this vector form is not generalized here for all $\mu,\nu$, but only for the $\textbf{q}$ part.\\
We are interested in studying the motion of particle as different parameters are varied slowly with time. For adiabatic evolution, the wave function changes slowly, and apart from a phase factor to the first order in the rate of change of the Hamiltonian, the wave function is given by: 
\begin{equation}
	\ket{\psi(n)}=\left | u_n \right > -i \hbar \sum_{n'\neq n}\frac{\left | u_{n'}\right>\left <u_{n'} | \frac{\partial u_{n'}}{\partial t}\right>}  {(\varepsilon_n-\varepsilon_{n'})}
\end{equation}

Now the velocity operator in the q-representation has the form $\textbf{v}(q,t)= \frac{1}{\hbar} \mathbf{\nabla}_q H(q,t)$, so the average velocity in a state of given q to the first order is given by
\begin{equation}
\begin{split}
	\textbf{v}_n(\textbf{q})&=\frac{1}{\hbar}\nabla_{\textbf{q}}\varepsilon_n(\textbf{q})\\&-i\sum_{n'\neq n}\left\lbrace\frac{\left< u_n\right | \nabla_\textbf{q} H(\textbf{q}) \left | u_{n'}\right >\left<u_{n'}|\frac{\partial u_n}{\partial t}\right >}{\varepsilon_n-\varepsilon_n'}-c.c\right\rbrace,
\end{split}
\end{equation}
where c.c denotes the complex conjugate. Using the fact that when, $n\neq n'$, $\left< u_n\right | \nabla_\textbf{q} H(\textbf{q}) \left | u_{n'}\right >= (\varepsilon_n-\varepsilon_{n'})\left<\nabla_{\textbf{q}} u_n |u_{n'}\right >$, and the identity $\sum \left |u_{n'}\right >\left <u_{n'}\right |=1$, we find,
\begin{equation}\label{eq: vq}
	\textbf{v}_n(\textbf{q})=\frac{1}{\hbar}\nabla_{\textbf{q}}\varepsilon_n(\textbf{q})-i\left[\left<\nabla_{\textbf{q}}u_n| \frac{\partial u_n}{\partial t}\right> -\left <\frac{\partial u_n}{\partial t} | \nabla_{\textbf{q}} u_n\right >\right ].
\end{equation}
The $t$ dependence is coming through $\textbf{R}$, and thus we can write
\begin{equation} \label{eq: gradq}
	\left| \frac{\partial u_n}{\partial t}\right>=\dot{\textbf{q}}\cdot\nabla_{\textbf{q} }\left | u_n\right > + \dot{J_\mu}\left | \frac{\partial u_n}{\partial J_\mu}\right >,
\end{equation}
where summation over $\mu$ is implied. Substituting this expression in Eq.~(\ref{eq: vq}), we get the $i$th component of velocity as:

\begin{equation}
	\begin{split}
		v_i(\textbf{q})=&\frac{\partial \varepsilon_n}{\hbar\partial q_i}-i \dot{q_j}\left[\left<\frac{\partial u_n}{\partial q_i}|\frac{\partial u_n}{\partial q_j}\right>- \left<\frac{\partial u_n}{\partial q_j}|\frac{\partial u_n}{\partial q_i}\right>\right ]\\&
		-i \dot{J_\mu}\left[\left<\frac{\partial u_n}{\partial q_i}|\frac{\partial u_n}{\partial J_\mu}\right>- \left<\frac{\partial u_n}{\partial J_\mu}|\frac{\partial u_n}{\partial q_i}\right>\right ].
	\end{split}
\end{equation}
Using Eqs.~(\ref{eq: Aq}), (\ref{BC2}) and (\ref{eq: vector form}), we get:
\begin{equation} 
	\begin{split}
		\label{eq: finalvq}
		v_i(\textbf{q})=&\frac{\partial \varepsilon_n(\textbf{q})}{\hbar\partial q_i}-\epsilon_{ijk}\dot{q_j}\mathbf{\Omega}^n_k \\&
		-i \dot{J_\mu}\left[\left<\frac{\partial u_n}{\partial q_i}|\frac{\partial u_n}{\partial J_\mu}\right>- \left<\frac{\partial u_n}{\partial J_\mu}|\frac{\partial u_n}{\partial q_i}\right>\right ].
	\end{split}
\end{equation}
The last term in  Eq.~(\ref{eq: finalvq}) above can be expanded as:
\begin{equation}
	\begin{split}
		i &\left[\left<\frac{\partial u_n}{\partial q_i}|\frac{\partial u_n}{\partial J_\mu}\right>- \left<\frac{\partial u_n}{\partial J_\mu}|\frac{\partial u_n}{\partial q_i}\right>\right ]\\&= -i \frac{\partial }{\partial J_\mu}\left<u_n|\frac{\partial}{\partial q_i}|u_n\right>
		+i\left<\frac{\partial u_n}{\partial q_i}|\frac{\partial u_n}{\partial J_{\mu}}\right>\\&+i\left<u_n|\frac{\partial}{\partial q_i}\frac{\partial}{\partial J_{\mu}}|u_n\right>
		= -\frac{\partial A_{q_i}^n}{\partial J_{\mu}}+i\frac{\partial}{\partial q_i}\left<u_n|\frac{\partial}{\partial J_{\mu}}|u_n\right>,
	\end{split}
\end{equation}
and thus the Eq.~(\ref{eq: finalvq}) becomes:
\begin{equation}
	\begin{split}
		\label{eq: tpart}
		v_i(\textbf{q})=&\frac{\partial \varepsilon_n(\textbf{q})}{\hbar\partial q_i}- (\dot{\textbf{q}}\times (\nabla \times \mathbf{A}^n))_i 
		+ \dot{J_\mu}\frac{\partial \textbf{A}_i^n}{\partial J_\mu}\\&-i\dot{J_{\mu}}\frac{\partial}{\partial q_i}\left<u_n|\frac{\partial}{\partial J_{\mu}}|u_n\right>.
	\end{split}
\end{equation}
It can  be written more concisely as:
\begin{equation}\label{eq:ffeq}
	\dot{r_i}(\textbf{q})=\frac{\partial \varepsilon_n(\textbf{q})}{\hbar\partial q_i}- (\dot{\textbf{q}}\times (\nabla \times \mathbf{A}^n))_i  
	+ \left(\frac{\partial \textbf{A}^n_{i}}{\partial t}\right)_{\textbf{q}}-\nabla_{\textbf{q}} \chi_n(t),
\end{equation}
where 
\begin{equation}
	\chi_n(t)=i\left<u_n|\frac{\partial}{\partial t}|u_n\right>.
\end{equation}
This equation has a striking similarity with the equation of motion of a charged  particle in the presence of an electric and magnetic field as given by:
\begin{equation}
	\hbar\dot{\textbf{k}}= Q\nabla_{\textbf{r}}\Phi+Q \dot{\textbf{r}}\times (\nabla_{\textbf{r}}\times \textbf{A})- Q\left(\frac{\partial \textbf{A}}{\partial t}\right)_\textbf{r},
\end{equation}
where $Q$ is the electric charge, $\Phi$  the scalar potential, and $\textbf{A}$ is the electromagnetic vector potential. This allows us to define analogs of electric and magnetic field from the geometric quantities in Eq.~(\ref{eq:ffeq}) as follows:
\begin{eqnarray}
	\textbf{B}&\rightarrow&\mathbf{\Omega}^n,\\
	\textbf{E}&\rightarrow&\left(\frac{\partial\textbf{A}^n}{\partial t}-\nabla_{\textbf{q}}\chi_n\right).
\end{eqnarray}

These quantities are gauge invariant as shown below.

\subsection{Gauge Invariance of the Electric field analog}
From electrodynamics, we know that $\textbf{E}= -\nabla_{\textbf{r}}(\Phi)-\frac{\partial \textbf{A}}{\partial t}$, and $\textbf{B}=\nabla\times\textbf{A}$ are gauge invariant quantities. So in this section, we prove that a similar gauge invariance is satisfied by their  analogs. 
If we make a gauge transformation 
\begin{equation}
	\left |u_n(\textbf{R})\right>\rightarrow e^{i\zeta(\textbf{R})}\left | u_n(\textbf{R}) \right >,
\end{equation}
where $ \zeta  (\textbf{R} )$ is an arbitrary smooth function, then $ \textbf{A}^n $ transforms as follows:
\begin{equation}
	\textbf{A}^n\rightarrow\textbf{A}^n-\nabla_{\textbf{q}}\zeta(\textbf{q,G}).
\end{equation}
In analogy to the EM vector potential we want to show that in Eq.~(\ref{eq:ffeq}), $\left(\frac{\partial \textbf{A}^n}{\partial t}\right)_{\textbf{q}}-\nabla_\textbf{q}\chi_n$ is a gauge invariant quantity. Let $\left |u_n'(\textbf{R})\right>=e^{i\zeta(\textbf{R})}\left |u_n(\textbf{R})\right>$, then the term $\nabla_\textbf{q}\chi$ transforms as :
\[\nabla_\textbf{q} \chi'=i\nabla_\textbf{q}\left<u_n'|\frac{\partial}{\partial t}|u_n'\right>=i\nabla_\textbf{q}\left[ \left<u_n|\frac{\partial}{\partial t}|u_n\right>+i\frac{\partial\zeta}{\partial t}\right],\]
\begin{equation}
	\nabla_\textbf{q} \chi'=\nabla_\textbf{q}\chi-\nabla_\textbf{q}\frac{\partial\zeta}{\partial t},
\end{equation}
and the other term $\left(\frac{\partial \textbf{A}'^n}{\partial t}\right)_{\textbf{q}}$  transforms as follows
\begin{equation}\label{eq: achange}
	\left(\frac{\partial \textbf{A}'^n}{\partial t}\right)_{\textbf{q}}= \left(\frac{\partial \textbf{A}^n}{\partial t}\right)_{\textbf{q}} -\nabla_{\textbf{q}}\left(\frac{\partial \zeta}{\partial t}\right)_{\textbf{q}}.
\end{equation}

So, the expression  $\left(\frac{\partial \textbf{A}^n}{\partial t}\right)_{\textbf{q}}-\nabla_\textbf{q}\chi_n$ is  gauge invariant, and hence the equation of motion is not modified under the gauge transformation. This term $\left(\frac{\partial \textbf{A}^n}{\partial t}\right)_{\textbf{q}}-\nabla_{\textbf{q}}\chi_n$ is analogous  to the electric field $\textbf{E}$ in the real space.

\section{Derivation for equation of motion for a Bloch wavepacket undergoing Rabi Oscillations}\label{appendixb}
Consider a wavepacket with support on two Bloch bands. We want to study its evolution under the  Hamiltonian  $H-\textbf{F}(t)\cdot\hat{\textbf{r}}$, where $H$ is translationally invariant, and $\textbf{F}(t)$ is the force applied on the wavepacket. The wavefunction describing the system is given by Schrodinger equation:
\begin{equation}
	i\frac{\partial}{\partial t}\ket{\Psi(\textbf{r},t)}=\left(H(t)-\textbf{F}(t)\cdot\hat{\textbf{r}}\right)\ket{\Psi(\textbf{r},t)}.
	\label{mainschrd}
\end{equation}
Here, we can express  $\ket{\Psi(\textbf{r},t)}$ as a superposition of  Bloch wavefunctions. For a two band model we can write:
\begin{equation}
	\ket{\Psi(\textbf{r},t)}=\sum_{n=1,2}\int_{BZ}d^2\textbf{k}\phi_n(\textbf{k},t)e^{i\textbf{k}\cdot\textbf{r}}\ket{u_n(\textbf{k})},
\end{equation}
where  $\ket{u_n(\textbf{k})}$ is the cell-periodic part of $n^{th}$ Bloch wavefunction.  Substituting in Eq.~(\ref{mainschrd}) above, we get:
\begin{equation}
	\begin{split}
		i\frac{\partial}{\partial t}(\phi_n&(\textbf{k},t))=\\\sum_{m=1,2}&\int_{BZ}d^2\textbf{k}(\phi_{m}(\textbf{k},t)\bra{u_n(\textbf{k},t)}{H\ket{u_{m}(\textbf{k},t)}}\\&-i\bra{u_{n}(\textbf{k},t)}\textbf{F}\cdot\nabla_{\textbf{k}}(\phi_{m}(\textbf{k},t)\ket{u_{m}(\textbf{k})})) .
	\end{split}
\end{equation}

Now, since $H$ is translationally invariant, and there is no mixing between  different $\textbf{k}$ components, so we can use the ansatz:

\begin{equation}
	\ket{\Psi(\textbf{r},t)}=\sum_{n=1,2}\int_{BZ}d^2\textbf{k}\phi_n(\textbf{k}(t),t)e^{i\textbf{k}(t)\cdot\textbf{r}}\ket{u_n(\textbf{k}(t))},
\end{equation}
where $\textbf{k}(t)=\textbf{k}+\int_{0}^{t}\textbf{F}(t')dt'$ and defining  :
\begin{equation}
	\psi_n(\textbf{k},t)\equiv\phi_n(\textbf{k}(t),t),
\end{equation} we get the following equation for $\psi_n(\textbf{k},t)$
\begin{equation}
	i\frac{\partial}{\partial t}\psi_n(\textbf{k},t)=\sum_{m}\left(H^{n,m}(t)-\textbf{F}\cdot\textbf{A}^{n,m}(\textbf{k}(t))\right)\psi_{m}(\textbf{k},t),
\end{equation}
where $\textbf{A}^{n,m}(\textbf{k})=i\bra{u_n(\textbf{k})}\nabla_{\textbf{k}}\ket{u_{m}(\textbf{k})}$. This ansatz was used mainly to  consider the fact that probability distribution in $\textbf{k}$ moves in time in the presence of a force, and the  coefficients at $\textbf{k}+\int_0^t\textbf{F}(t)dt$ are decided by the initial conditions at $\textbf{k}$. In the absence of an external force, we  can write :
\begin{equation}
	\left|\Psi(\textbf{r},t)\right>=\sum_{n=1,2}\int_{BZ} d^2\textbf{k}e^{i\textbf{k}\cdot\textbf{r}}\psi_n(\textbf{k},t)\ket{u_n(\textbf{k})}.
\end{equation}
For translationally invariant  $H$, there is no mixing between different $\textbf{k}$ and at each $\textbf{k}$, we have:
\begin{equation}
	i\frac{\partial}{\partial t}\psi_n(\textbf{k},t)=\sum_{m=1,2}H(\textbf{k},t)^{n,m}\psi_{m}(\textbf{k}),
\end{equation}
and  $|\psi_1(\textbf{k},t)|^2+|\psi_2(\textbf{k},t)|^2$ is a  function of $\textbf{k}$ only. This allows us to  express:
\begin{equation}
	\left|\Psi(\textbf{r},t)\right>=\int d^2\textbf{k}e^{i\textbf{k}\cdot\textbf{r}}\phi(\textbf{k})\ket{\Phi(\textbf{k},t)},
\end{equation}
where  $\phi(\textbf{k})=\sqrt{|\psi_1(\textbf{k},t)|^2+|\psi_2(\textbf{k},t)|^2}$ and  $\ket{\Phi(\textbf{k},t)}$ is a superposition of $\ket{u_n(\textbf{k})} $ with time dependent coefficients  such that 
\begin{equation}
	i\frac{\partial}{\partial t}\ket{\Phi(\textbf{k},t)}=H(\textbf{k},t)\ket{\Phi(\textbf{k},t)},
\end{equation}
and $\bra{\Phi(\textbf{k},t)}\ket{\Phi(\textbf{k},t)}=1$.

Now, we can do something similar in the presence of an external force by defining 
\begin{equation}
	\ket{\Psi(\textbf{r},t)}=\int d^2\textbf{k}e^{i\textbf{k}(t)\cdot\textbf{r}}\phi(\textbf{k})\ket{\Phi(\textbf{k}(t),t)},
\end{equation}
where $\phi(\textbf{k})=\sqrt{\phi_1(\textbf{k},0)^2+\phi_2(\textbf{k},0)^2}$, $\textbf{k}(t)=\textbf{k}+\int_0^t\textbf{F}(t')dt'$, and $\ket{\Phi(\textbf{k}(t),t)}$ is governed by $H-\textbf{F}\cdot\textbf{A}$. For the special case of sinusoidal force $\textbf{F}(t)=\textbf{F}_0\sin(\omega t)$, we notice that $\textbf{k}(t=nT)=\textbf{k}$, and hence at any time $t$ which is integer multiple of time period $T$, we can write:
\begin{equation}
	\ket{\Psi(\textbf{r},t)}=\int d^2\textbf{k}e^{i\textbf{k}.\textbf{r}}\phi(\textbf{k})\ket{\Phi(\textbf{k},t)},
\end{equation}
where $\ket{\Phi(\textbf{k},t)}$ is governed by:
\begin{equation}
	i\frac{\partial}{\partial t}\ket{\Phi(\textbf{k},t)}=\left(H_0-\textbf{F}(t)\cdot\textbf{A}(\textbf{k})\right)\ket{\Phi(\textbf{k},t)}.
	\label{F_H_appendixB}
\end{equation}
\subsection{Expression for velocity of C.O.M of wavepacket in presence of near resonant periodic drive}
For a two-level system, consider the full Hamiltonian
\begin{equation}
	H=H_0+H'=\begin{bmatrix}
		E_1& 0\\
		0 & E_2
	\end{bmatrix} +\begin{bmatrix}
		0& \frac{V}{2}e^{i\theta}e^{i\omega t}\\
		\frac{V}{2}e^{-i\theta}e^{-i\omega t} & 0
	\end{bmatrix},
	\label{hn}
\end{equation}
where $V$ is taken as real.
This gives 
\begin{equation}
	\ket{\Phi(\textbf{q},t)}=a(t)e^{i\omega t/2}\ket{u_1(\textbf{q},t)}+b(t)e^{-i\omega t/2}\ket{u_2(\textbf{q},t)},
	\label{phin}
\end{equation} and substituting in Eq.~(\ref{rr1}), and neglecting the terms oscillating at frequency $\omega$, we get
\begin{equation}
	\begin{split}
		\avg{\textbf{r}}&=\int d^2\textbf{q}|\phi(\textbf{q},\textbf{q}_0)|^2\bra{\Phi(\textbf{q},t)}\nabla_{\textbf{q}}\ket{\Phi(\textbf{q},t)}\\&
		=\avg{\textbf{A}_{nn}}|_{\textbf{q}_0}+a^*(t)\nabla_\textbf{q}a(t)+b^*(t)\nabla_\textbf{q}b(t).
		\label{rappendix}
	\end{split}
\end{equation}
Now, subsituting Eq.~(\ref{phin}) in Eq.~(\ref{hn}), we get
\begin{equation}
	\frac{\partial}{\partial t}a(t)=-i\left((E_1+\frac{\omega}{2})a(t)+\frac{V}{2}e^{i\theta}b(t)\right)
\end{equation}
and
\begin{equation}
	\frac{\partial}{\partial t}(b(t))=-i\left(\left(E_2-\frac{\omega}{2}\right)b(t)+\frac{V}{2}e^{-i\theta}a(t)\right).
\end{equation}
If the wavepacket starts in one of the eigen states, we get the following expression for velocity:
\begin{equation}
\begin{split}
	\textbf{v}=&|(a(t)|^2\nabla_\textbf{q}E_1+|b(t)|^2\nabla_\textbf{q}E_2\\&+\frac{\partial}{\partial t}\avg{\textbf{A}_{nn}}+\frac{\partial}{\partial t}\left(|a(t)|^2\nabla_\textbf{q}(\theta)\right).
\end{split}
\end{equation}
Given the fact that 
\begin{equation}
	\frac{\partial}{\partial t}|a(t)|^2=-\frac{\partial}{\partial t}|b(t)|^2,
\end{equation}
we can write 
\begin{equation}
	\textbf{v}=\avg{\nabla_\textbf{q}E_n}+\frac{\partial}{\partial t}\avg{\textbf{A}_{nn}}+\frac{\partial}{\partial t}\avg{\phi_n},
\end{equation}
where 
\begin{equation}
	\phi_1=-\phi_2=\frac{\theta}{2}=\frac{1}{2}\text{Arg}\bra{u_1}H'\ket{u_2}.
\end{equation}
It is worth mentioning again that the above equation does not take into account the fast oscillatory motion at frequency $\omega$. Now, if we include such terms then, Eq.~(\ref{rappendix}) is modified as follows:
\begin{equation}
	\begin{split}
		\label{oscillations}
		\avg{\textbf{r}}=&
		\avg{\textbf{A}_{nn}}|_{\textbf{q}_0}+a^*(t)\nabla_\textbf{q}(a(t))+b^*(t)\nabla_\textbf{q}b(t)\\&+a^*(t)b(t)e^{-i\omega t}\textbf{A}_{ge}+a(t)b^*(t)e^{i\omega t}\textbf{A}_{eg}.
	\end{split}
\end{equation}
\section{Results for different sizes of wavepacket}
\label{different_sigma}

Here, we consider the effects of increasing the size of wavepacket in quasi-momentum space. Since, the resonance condition is satisfied only at the center of wavepacket, so a larger wavepacket would have significant detuning. This detuning would decrease the transition probability after a Rabi cycle, and would also decrease the transverse drift observed during the  process of band switching. We plot the resulting transverse drift and excitation probabilities for three different wavepacket sizes, and other parameters are same as that for Fig.~\ref{Theta2} in the main text.
\begin{figure}
	\includegraphics[scale=0.5]{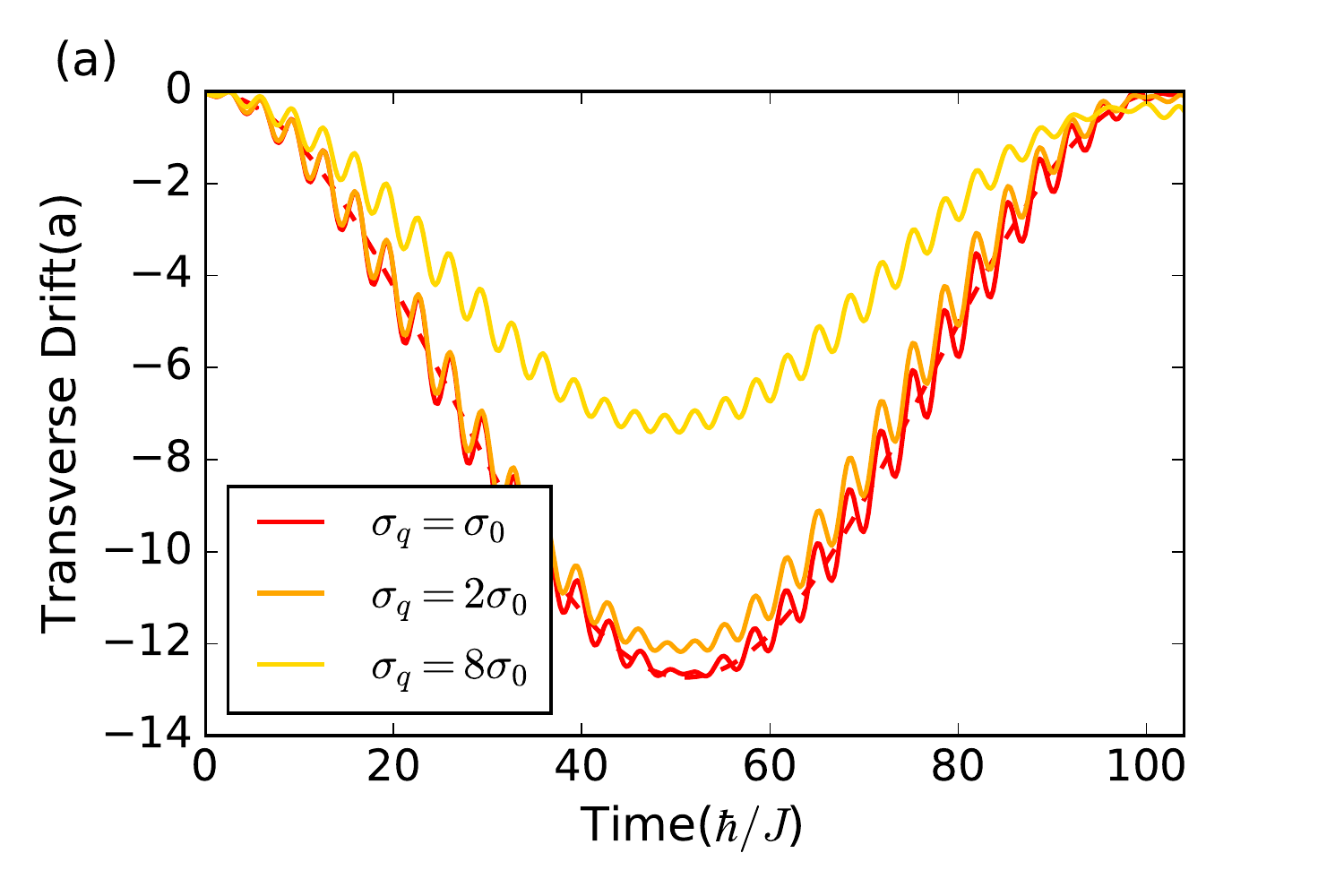}
	\includegraphics[scale=0.5]{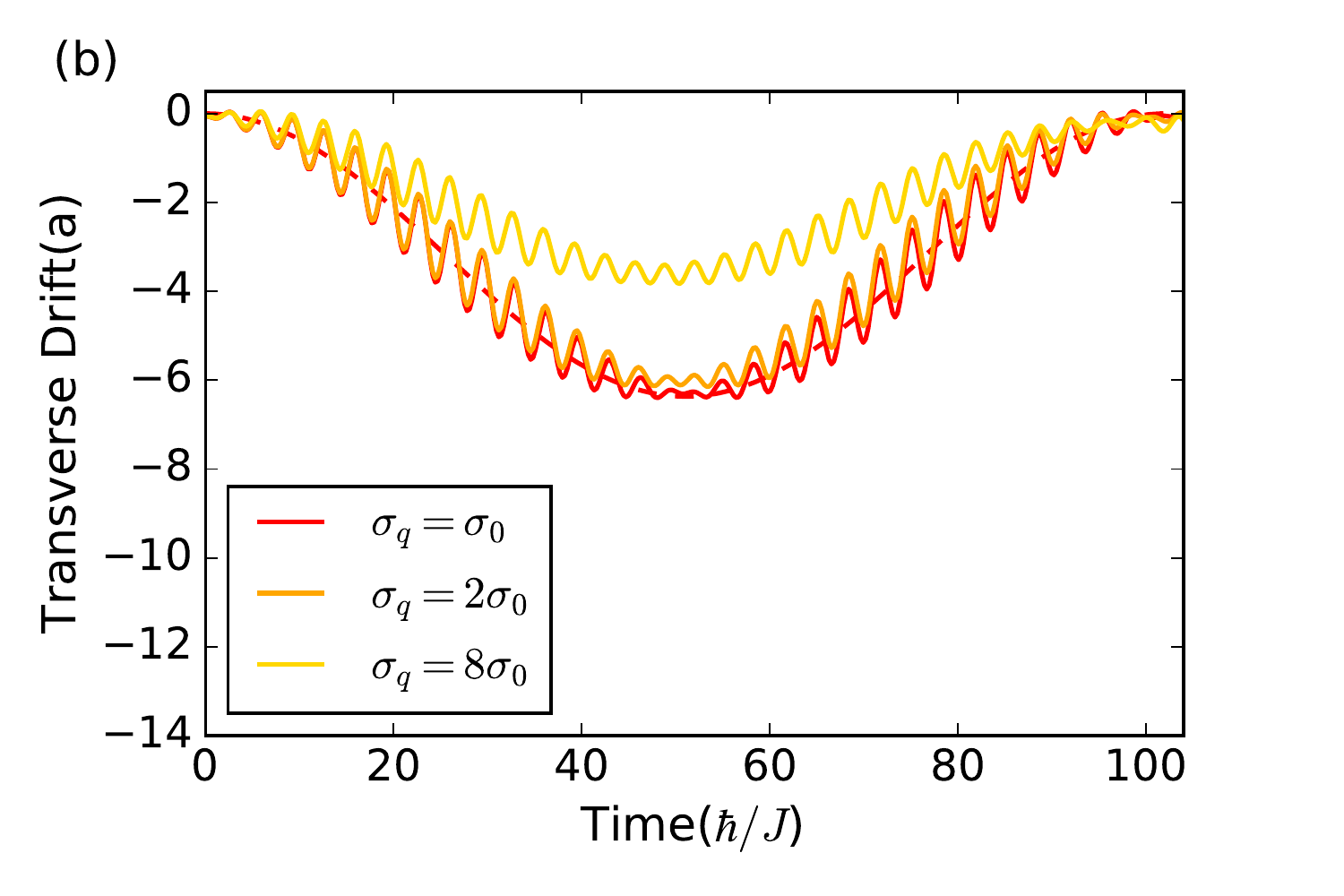}
	\includegraphics[scale=0.5]{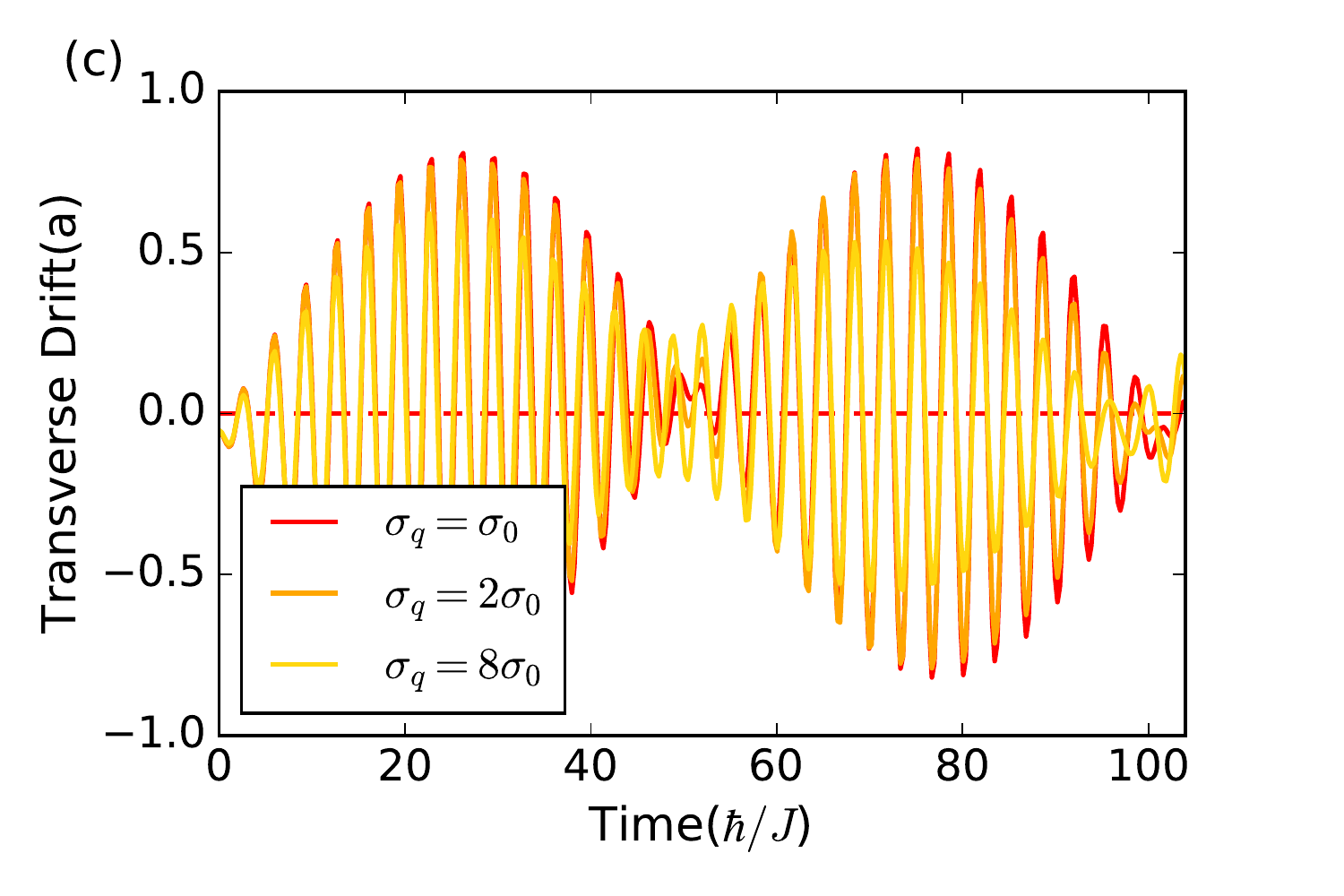}
	\includegraphics[scale=0.5]{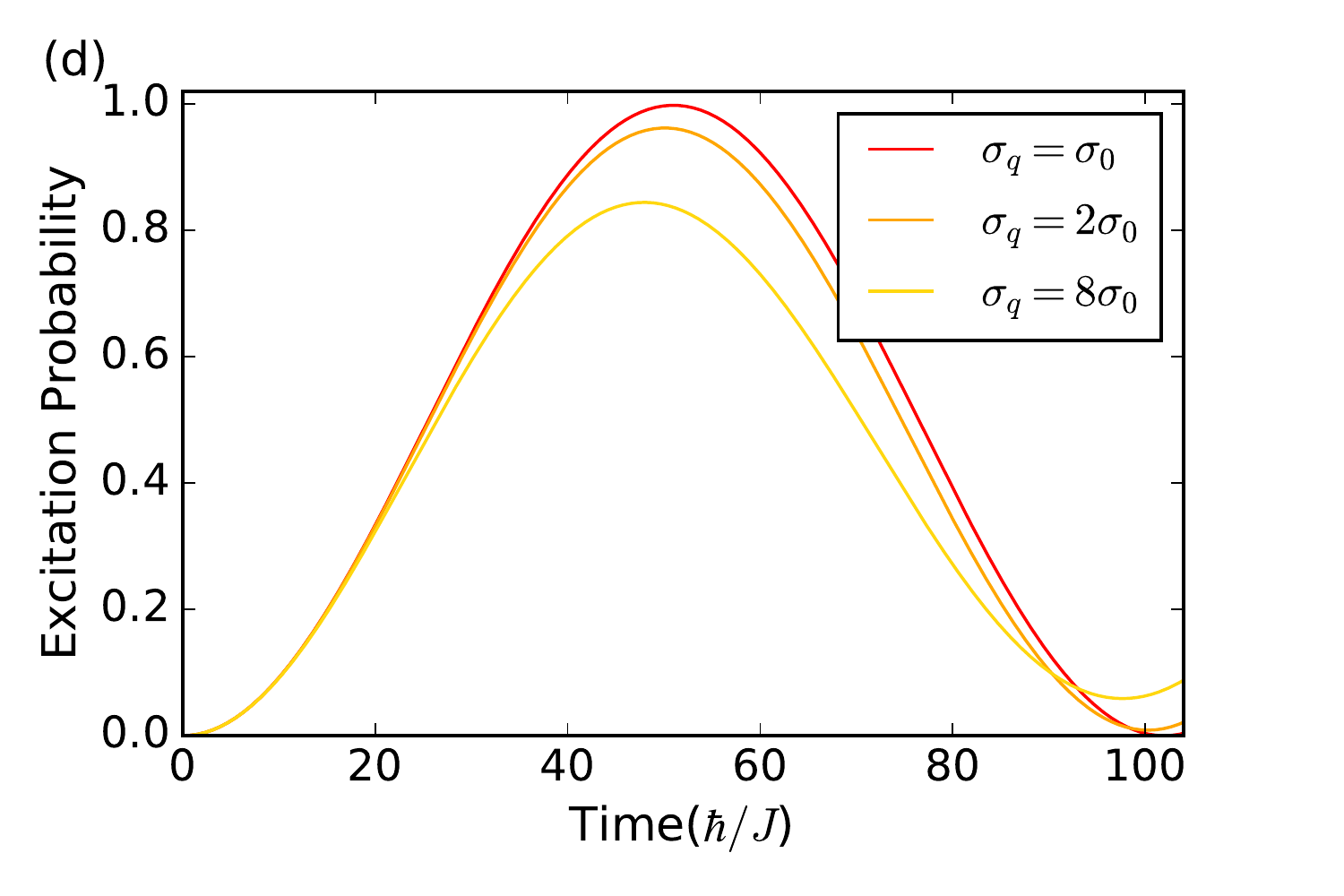}
	\caption{Transverse drift for different wavepacket size for (a)$\Theta=0.0$, (b)$\Theta=\phi/2$, (c)$\Theta=\phi$. Transition probability is same for all of the above cases and its dependence on wavepacket size is shown in (d). Here $\sigma_0$ is the size of wavepacket considered in the main text (Fig.~\ref{Theta2}).}
	\label{differentsgma_graph}
\end{figure}
\section{Pump from non-adiabatic processes only}\label{nonadiabaticpump}

\begin{figure}
	\includegraphics[scale=0.25]{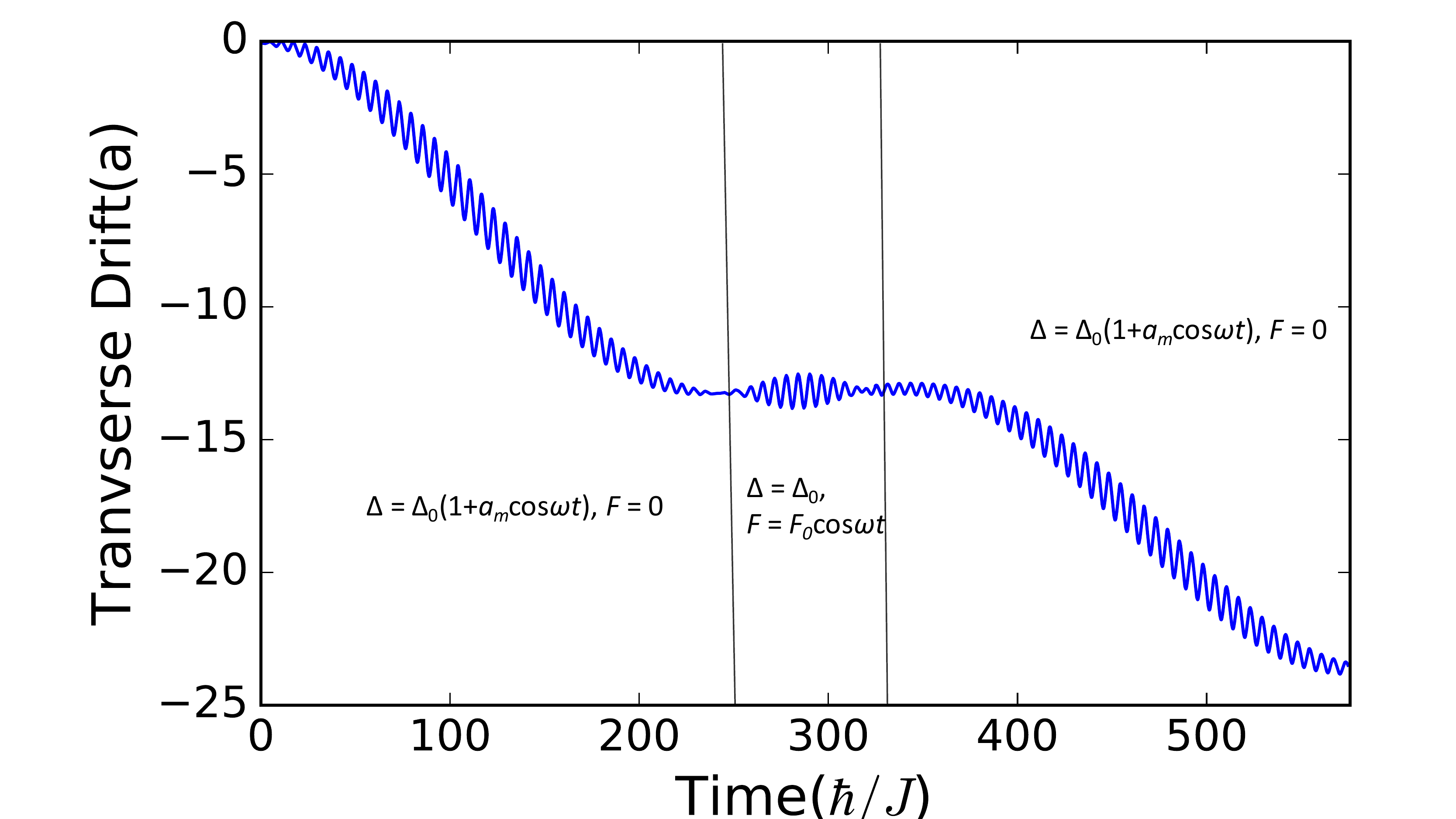}
	\caption{Transverse drift during three different band switching steps. For the first and third  step, we modulate the sublattice offset-energy $\Delta$, and the wavepacket moves from lower to upper band and observe a significant transverse drift. During the second step, we keep $\Delta$ constant, and apply a sinusoidal force and thus a negligible anomalous drift as expected. }
	\label{non_adiabatic_prob}
\end{figure}

In Sec.~\ref{pump}, we showed how one can combine adiabatic and non-adiabatic steps to amplify the anomalous drift obtained from the electric field analog. We mainly exploited the fact that shift vector was vanishingly small when transition was achieved using a time-periodic force, and the transverse drift was same during two adiabatic steps of the cycle. In Sec.~\ref{nonadiabatic}, we also showed that the shift vector is significantly large when band transition is achieved using the sublattice offset-energy modulation. Now, one can in fact combine these two different kind of non-adiabatic processes to amplify the transverse drift arising purely from the changes in the  Berry connection. Here, we simulate the motion of a wavepacket undergoing Rabi oscillations such that transition from lower to upper band is achieved by modulating offset energy, and use sinusoidal force for the opposite step. The wavepacket is shifted significantly during the sublattice offset-energy modulation step only, and its direction depends only on the bands involved in transition as shown in Fig.~\ref{non_adiabatic_prob}.
\section*{}
\begin{figure}
	\includegraphics[scale=0.47]{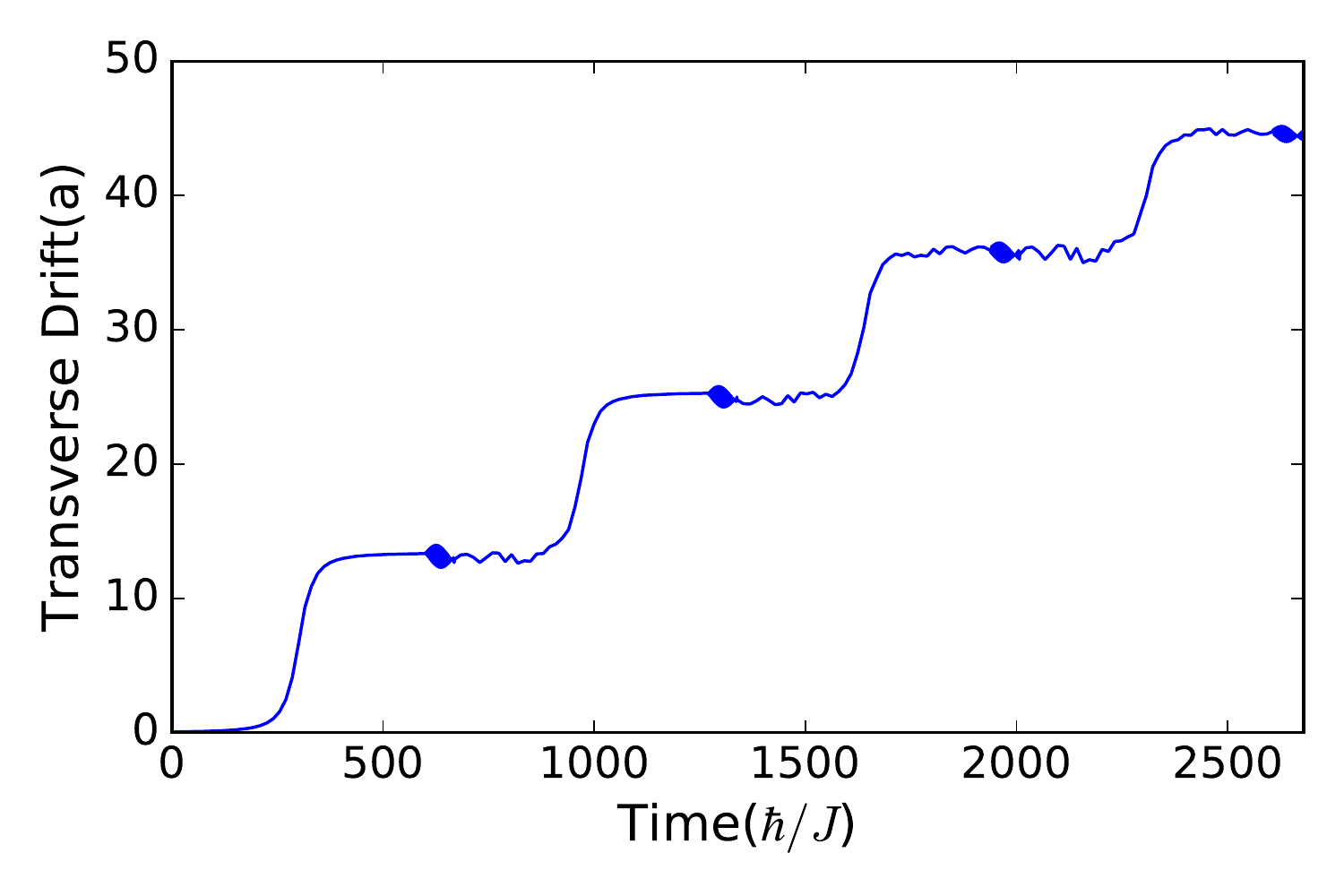}
	\caption{Transverse drift after two cycles of the scheme discussed in Sec.~\ref{pump}.}
	\label{twocycle}
\end{figure}
\newpage
\bibliographystyle{apsrev4-1}   
\bibliography{berry2.bib}
\end{document}